\documentclass[]{aa}
\usepackage{graphicx}
\usepackage{natbib}
\usepackage{amssymb}
\usepackage{txfonts}
\usepackage[figuresright]{rotating}

\bibpunct{(}{)}{;}{a}{}{,}
\newcommand{\be}{\begin{equation}}
\newcommand{\ee}{\end{equation}}
\newcommand{\phn}{\phantom{0}}

\begin{document}

\title{Polarization microvariability of BL\,Lac objects
\thanks{Based on observations made at the Complejo
Astron\'omico El Leoncito, which is operated under agreement
between CONICET and the National Universities of La Plata,
C\'ordoba, and San Juan.}
\author{I. Andruchow\inst{1,2} \thanks{Fellow of CONICET},
G. E. Romero\inst{1,2} \thanks{Member of CONICET}, S. A.
Cellone\inst{2} $^{\star \star \star}$} } \institute{Instituto
Argentino de Radioastronom\'{\i}a, C.C.5, (1894) Villa Elisa,
Buenos Aires, Argentina \and Facultad de Ciencias Astron\'omicas y
Geof\'{\i}sicas UNLP, Paseo del Bosque, B1900FWA La Plata,
Argentina}

\offprints{}
\date{Received / Accepted}

\titlerunning{Polarization microvariability of BL\,Lac objects}
\authorrunning{I. Andruchow et al.}

\abstract{We present the results of a systematic observational campaign
designed to search for microvariability in the optical polarization of BL
Lac objects. We have observed a sample formed by 8 X-ray-selected and 10
radio-selected sources, looking for rapid changes in both the degree of
linear polarization and the corresponding polarization angle. The whole
campaign was carried out along the last three years, and most of the objects
were observed at least on two consecutive nights.  The statistical
properties of both classes of BL\,Lac objects are compared, and some general
conclusions on the nature of the phenomenon are drawn. In general, radio
selected sources seem to display higher duty cycles for polarimetric
microvariability and, on average, they have a stronger polarization.

\keywords{Galaxies: active
--- BL Lacertae objects: general --- polarization}}

\maketitle

\section{Introduction}

The existence of rapid changes (time scales from minutes to hours)
in the optical brightness of blazars is a well established fact
\citep{R70, MCG89, RCC99, SGSW04, SGG05}.  These variations are
usually called \emph{microvariability} or \emph{intranight}
variability. The incidence of the phenomenon on different classes
of active galactic nuclei (AGNs) seems to be also different
\citep{JM97, RCC99, RCCA02, SGG05}.  BL\,Lacertae objects, along
with flat radio-spectrum quasars, seem to be among the most
variable sources on short time scales. BL\,Lacs, in turn, can be
divided into two groups: X-ray selected BL\,Lacs (XBLs) and
radio-selected BL\,Lacs (RBLs) according to their spectral energy
distributions (SEDs). In general, both types of objects have SEDs
with two peaks, one thought to be due to synchrotron emission and
the other produced by inverse Compton upscattering of lower energy
photons. In the case of XBLs, the synchrotron peak falls in the
X-ray band. In RBLs this peak is shifted towards lower energies,
being between the radio and infrared bands. The optical
microvariability behaviour of both sub-types of BL\,Lacs seems to
be quite different \citep{HW96, HW98, RCCA02}.  The XBLs are
usually less variable, with smaller duty cycles and variability
amplitudes than the RBLs.

Although the optical microvariability phenomenon has been extensively
studied for different AGNs, little is known about the polarimetric behaviour
of these objects on time scales of hours. High-temporal resolution
polarimetry has been performed for a few particular sources
\citep[e.g. 3C279,][]{ACR03} confirming the existence of microvariability
in the optical polarization of some objects, but there is still no
statistically significant information.

In this paper we present, by first time, results of a systematic search for
microvariability in the polarization of a sample of BL\,Lac objects. We have
looked for rapid changes in the degree of linear polarization and in the
position angle of both XBLs and RBLs. Our total sample has 18 objects, so we
will be able to extract some statistical conclusions on the duty cycles for
these sub-types of sources. For most of the objects, this is the first time
that high time resolution polarimetric observations are performed.

The structure of the paper is as follows. In Section 2 we present a detail
of the observed sample.  In Section 3 we describe the polarimetric
observations. In Section 4.1 we describe the statistical data
analysis. Section 4.2 presents individual notes on the behaviour of each
source. In Section 4.3 we briefly comment on the normalized Stokes
parameters. In Section 5 we discuss the statistics and the origin of the
observed variability. Finally, we close with our conclusions in Section 6.

\section{Sample\label{s_sam}}

The sample adopted for this work consists of 18 sources: 10
radio-selected blazars (RBL) and 8 X-ray-selected blazars (XBL),
and is taken from the AGN catalogs by \citet{V98} and
\citet{PG95}. We selected blazars with declinations lower than
$+20^{\circ}$ and brighter (at the time of our observations) than
magnitude $V = 18.5$. The redshifts spanned the range from $z =
0.044$ to $z = 1.048$. Basic general information on these objects
is given in Table \ref{tab1}. In this table, Column (1) gives the
name of the sources, Columns (2) and (3) give their equatorial
coordinates, Column (4) their classification, Columns (5), (6) and
(7) provide the colour excesses, the redshifts and the visual
magnitudes taken from the NASA Extragalactic Database (NED),
Column (8) gives the published maximum degree of optical linear
polarization (for those cases in which this value was know from
the literature), and Column (9) gives the corresponding references
for Column 8.

\begin{table*}
\begin{minipage}[t]{\hsize}
\renewcommand{\footnoterule}{}
\caption{Observed sample: general information}
\label{tab1} \centering
\begin{tabular}{@{}lcccccccccc}
\hline\noalign{\vskip 2pt} \hline \noalign{\smallskip}
 Object & $\alpha$(2000) & $\delta$(2000) & Type & $E(B-V)$ & $z$ & $\langle
m_{v}\rangle$ & $P_\mathrm{max}\pm \sigma$ & Ref.\footnote{IT:
 \citet{IT90}; SF: \citet{SF97}; F: J.H. Fan, private communication} \\
 & [h m s] & [$^{\circ} ~~ ' ~~ ''$] & & & & [mag] & [\%] & \\
\noalign {\smallskip} \hline
\noalign{\smallskip}
0118$-$272 & 01:20:31.6 & $-27$:01:25 & RBL & 0.014 & 0.559 & 16.5 &
$18.7\pm{1.0}$ & SF\\
0422+004 & 04:24:46.8 & $+00$:36:06 & RBL & 0.101 & 0.310 & 17.0 & $23.3\pm{1.1}$
& SF\\
0521$-$365 & 05:22:58.0 & $-36$:27:31 & RBL & 0.039 & 0.055 & 14.5  &
$\phn6.0\pm{1.5}$ & SF\\
0537$-$441 & 05:38:50.3 & $-44$:05:09 & RBL & 0.038 & 0.894 & 15.5  &
$18.7\pm{0.5}$ & SF\\
0548$-$322 & 05:50:39.7 & $-32$:16:18 & XBL & 0.035 & 0.069 & 18.3 &
$\phn1.4\pm{0.8}$ & SF\\
0558$-$385 & 06:00:22.1 & $-38$:53:55 & XBL & 0.054 & 0.044 &  ---  &  --- & F \\
0829+046 & 08:31:48.9 & $+04$:29:39 & RBL & 0.033 & 0.180 & 16.5  & $12.0\pm{3.0}$
& SF\\
1026$-$174 & 10:26:58.5 & $-17$:48:58 & XBL & 0.058 & 0.114 & 16.6  &  --- & \\
1101$-$232 & 11:03:37.6 & $-23$:29:30 & XBL & 0.059 & 0.186 & 16.6  &
$\phn7.4\pm{1.5}$ & SF\\
1144$-$379 & 11:47:01.4 & $-38$:12:11 & RBL & 0.096 & 1.048 & 16.2  &
$\phn8.5\pm{1.7}$ & IT\\
1312$-$422 & 13:15:03.4 & $-42$:36:50 & XBL & 0.105 & 0.108 & 16.6  &  --- & \\
1440+122 & 14:42:48.2 & $+12$:00:40 & XBL & 0.028 & 0.162 & 17.1  &  --- & \\
1510$-$089 & 15:12:50.5 & $-09$:06:00 & RBL & 0.097 & 0.360 & 16.5  &
$\phn1.9\pm{0.4}$ & SF\\
1514$-$241 & 15:17:41.8 & $-24$:22:19 & RBL & 0.138 & 0.049 & 15.1  &
$\phn6.9\pm{1.3}$ & SF\\
1553+113 & 15:55:43.0 & $+11$:11:24 & XBL & 0.052 & 0.360 & 15.0  &  --- & \\
1749+093 & 17:51:32.8 & $+09$:39:01 & RBL & 0.180 & 0.322 & 16.8  & $31.3\pm{0.6}$
& SF\\
2005$-$489 & 20:09:25.4 & $-48$:49:54 & RBL & 0.056 & 0.071 & 15.3  &
$\phn2.0\pm{0.2}$ & SF\\
2155$-$304 & 21:58:52.0 & $-30$:13:32 & XBL & 0.022 & 0.116 & 14.0  &
$10.3\pm{0.3}$ & SF\\
\noalign{\smallskip} \hline
\end{tabular}
\end{minipage}
\end{table*}

\section{Polarimetric observations and data reduction\label{s_obs}}

The observations were done with the 2.15-m Jorge Sahade telescope
at CASLEO, San Juan, Argentina, during 22 nights in April and
November 2002, May 2003, and April 2004. In all occasions, we used
the CASPROF photopolarimeter. This is an instrument developed at
CASLEO and based on other, similar two-channel photopolarimeters,
as MINIPOL and VATPOL \citep{M84,Martinez}. The observations were
carried out using always the same configuration: a Johnson $V$
filter and an 11.3 arcsec aperture diaphragm. Integration times
varied between 300 and 900~s, depending on the object brightness
and the quality of the night. In all cases, we observed the target
followed by a sky integration. Standard stars chosen from the
catalog by \citet{T90} were observed to determine the zero point
for the position angle and the instrumental polarization; the
latter was found to be practically zero. Weather conditions varied
along the whole campaign, from photometric to partially cloudy
(thin cirrus).

The data were processed using a systematic method, after discarding some
data points affected by moonlight contamination or passing clouds.  We
averaged each two consecutive target observations (on the $Q - U$ plane) in
order to improve the signal-to-noise ratio.  A factor that affects both the
object and the sky, is the presence of the Moon above the horizon. However,
any systematic error, leading to spurious variations in the sky
polarization, should be removed when the data are reduced, because each sky
observation is made near in time and position to the corresponding source
measurement. To prevent against errors due to rapid sky variations, we
interpolated the sky flux and polarization (on the $Q - U$ plane) to the
time corresponding to each object observation, thus giving a more accurate
sky subtraction.

\begin{figure*}
\includegraphics[width=0.45\hsize]{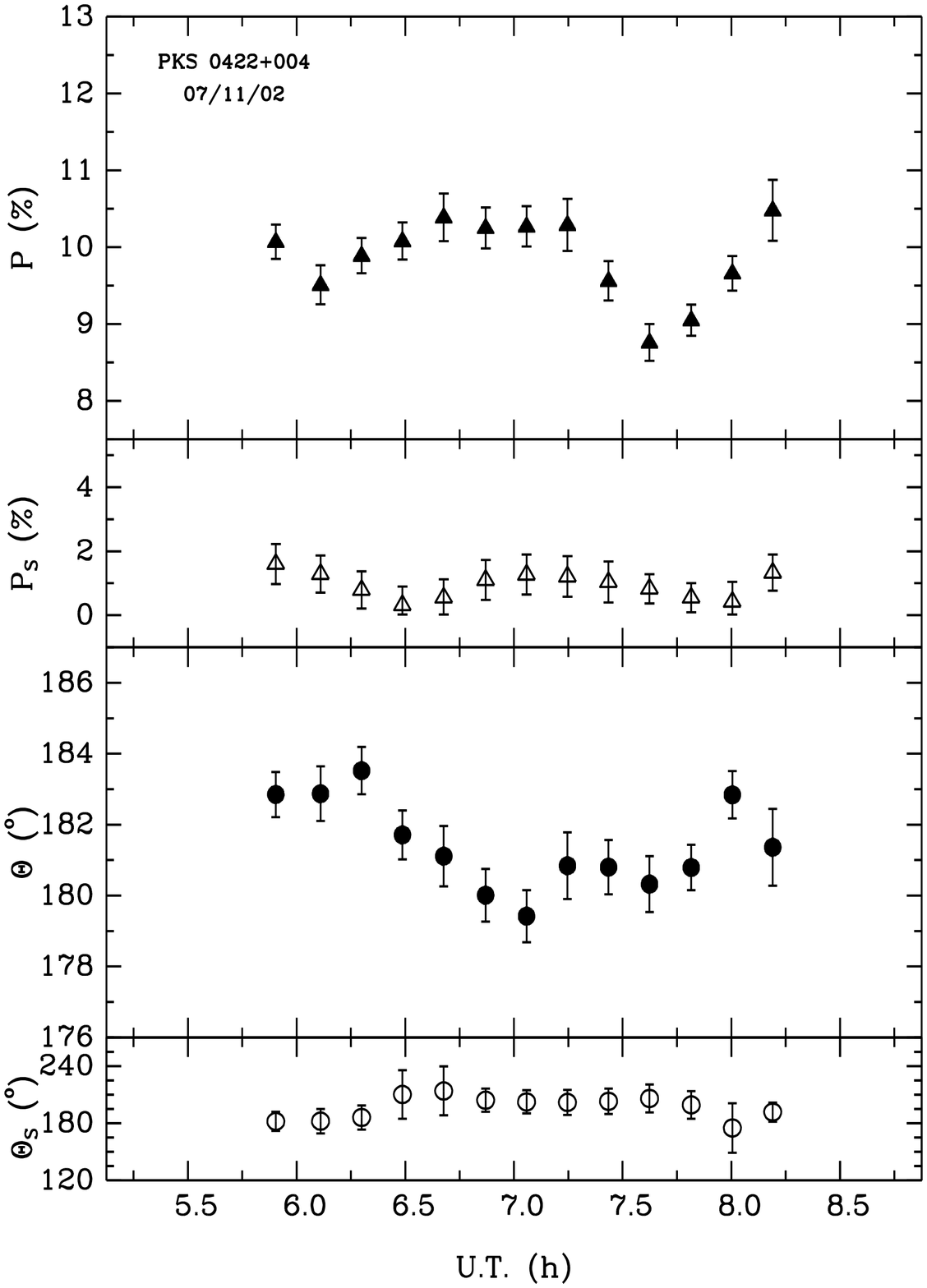} \qquad
\includegraphics[width=0.45\hsize]{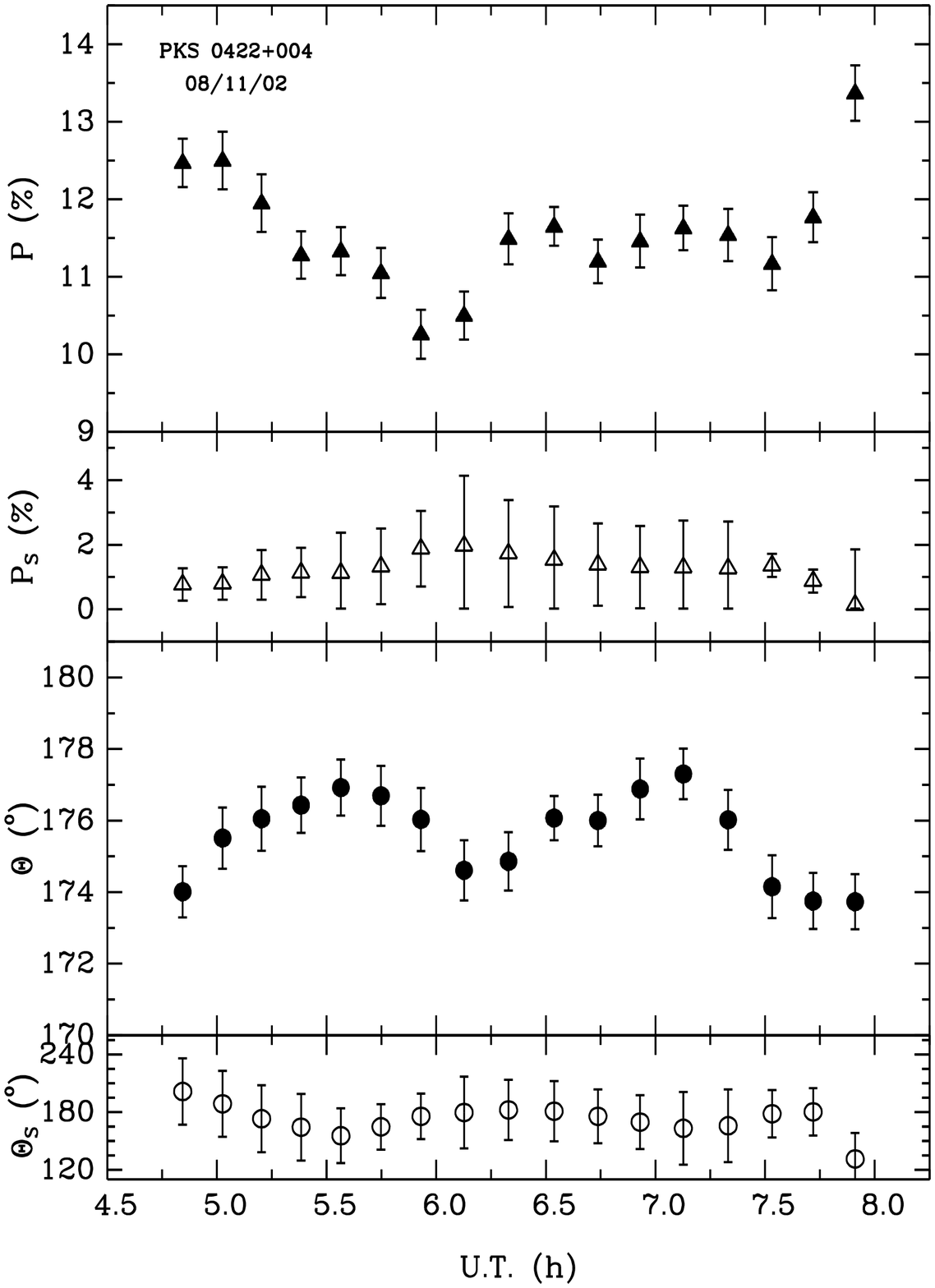}
\caption{Polarization and position angle for the RBL
PKS\,0422$+$004 (solid symbols) and for the corresponding sky
measurements (open
 symbols) as a function of time for two consecutive nights in November
 2002.} \label{0422}
\end{figure*}

\begin{figure*}
\includegraphics[width=0.45\hsize]{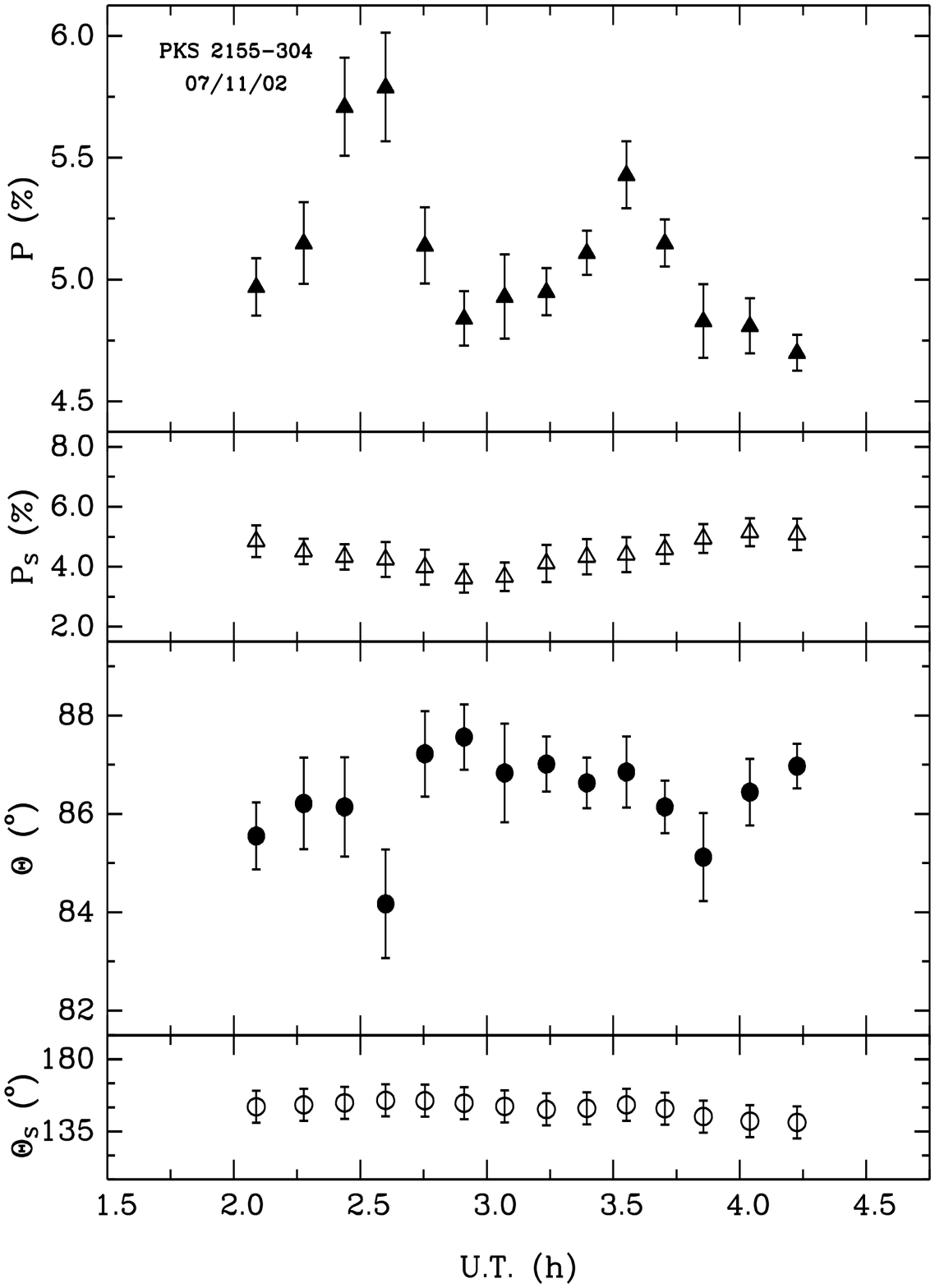} \qquad
\includegraphics[width=0.45\hsize]{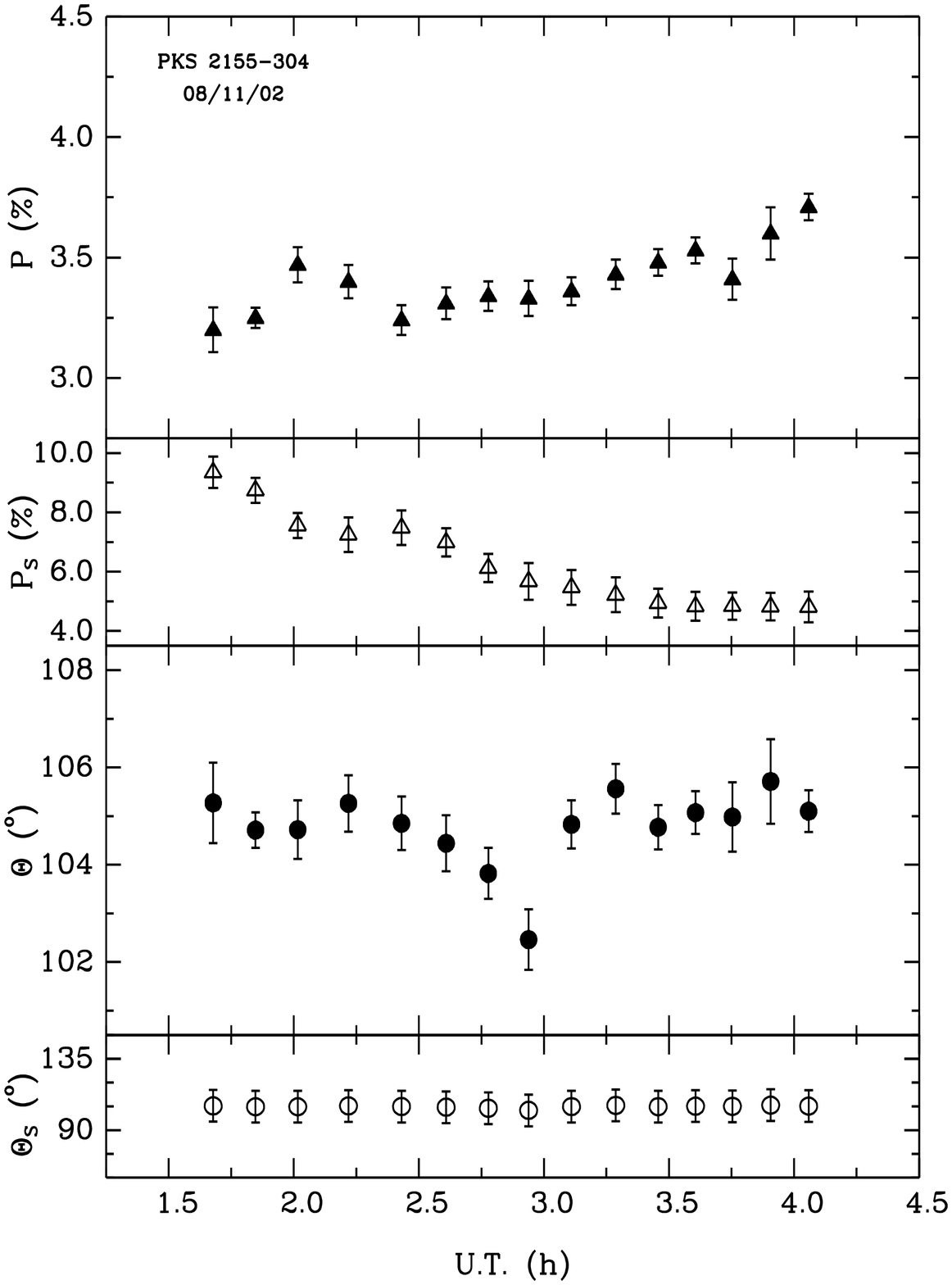}\\[10pt]
\includegraphics[width=0.45\hsize]{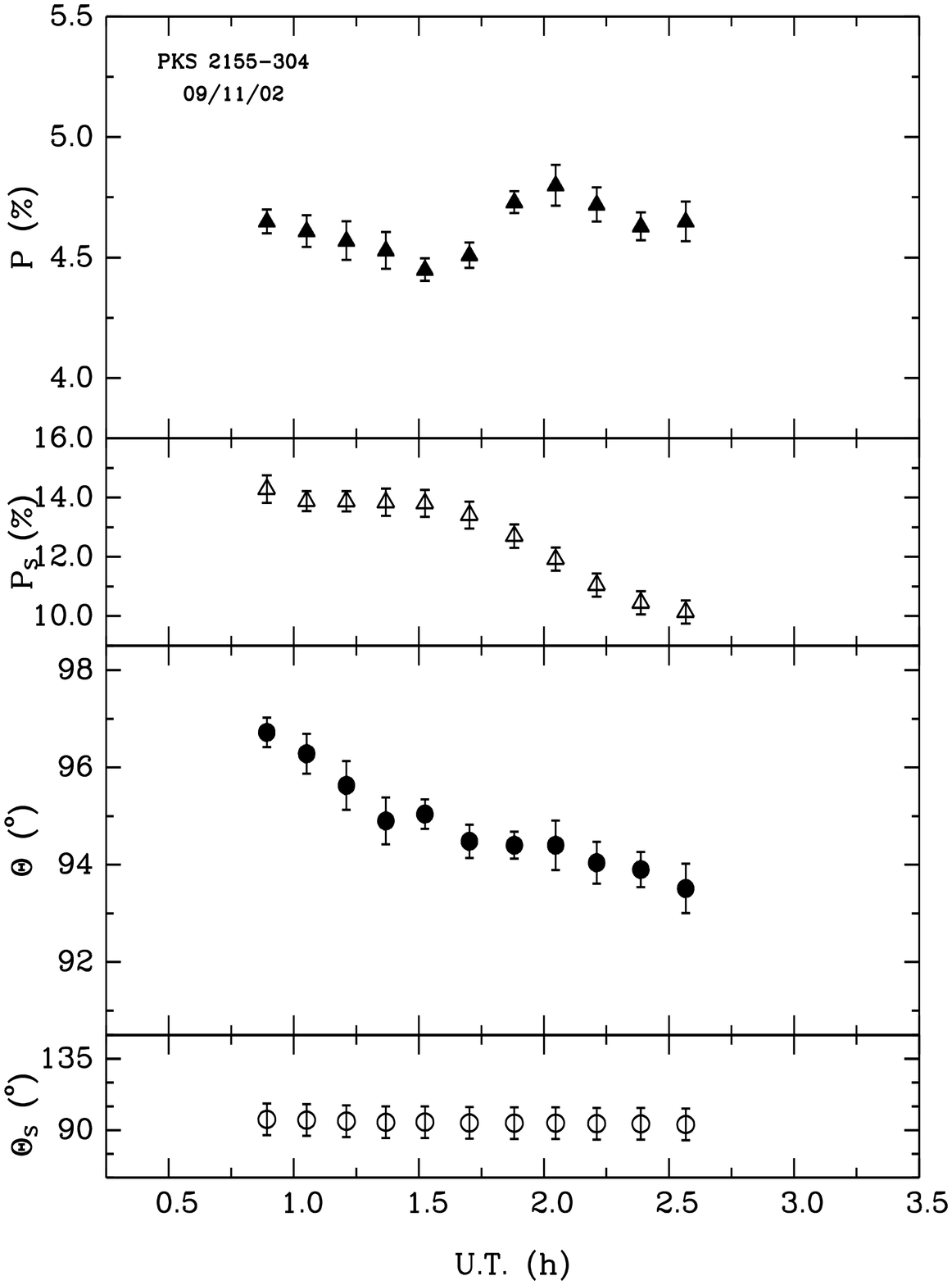}
\caption{Polarization and position angle for the XBL
PKS\,2155$-$304 (solid symbols) and for the corresponding sky
measurements (open
 symbols) as a function of time for three consecutive nights in November
 2002.}
 \label{2155}
\end{figure*}

For each observing session we searched for any residual systematic
errors by plotting the sky magnitude, as well as its polarization
percentage and position angle, against time, and then comparing
these graphs with the corresponding time-curves for the sources.
No spurious variations, due to rapid sky changes, were evident.
Two examples of typical microvariability curves are shown in
Figures~\ref{0422} and \ref{2155}; Fig.~\ref{0422} presents the
behaviour of the RBL object PKS\,0422$+$004 during two consecutive
nights in November 2002, whereas Fig.~\ref{2155} shows the
behaviour of the XBL source PKS\,2155$-$304 for three nights in
November 2002. Sky values are given as open symbols in both
figures; note that a different scale was used for the sky plots.
Both objects were always $\sim 0.5 - 3$ mag brighter than the sky,
thus confirming that any effect due to sky variations should not
be important.

We also checked for the incidence of the foreground polarization
(this is the polarization generated by interstellar dust particles
oriented by the magnetic field of the Galaxy). Following
\citet{H96}, we used the known relation $P_\mathrm{max}\, (\%) \le
9\, E_{(B-V)}$ to set an upper limit to the foreground
polarization in each of our fields. In column 5 of Table
\ref{tab1}\ we give the $E_{(B-V)}$ values taken from the NED; it
results that, since we observed at relatively high Galactic
latitudes, the values of $P_\mathrm{max}$ are between $0.13\%$ and
$1.62\%$.
 With the same purpose, we observed faint stars in most of the target
fields in order to check their polarization values. In all the cases, values
were $\ll 1 \%$. Thus, we are able to confirm that the foreground
polarization does not affect ours results.

We have used the Stokes parameters to check whether the pattern of
the observations was random in the $Q-U$ plane or not. We
consider, as usual, normalized dimensionless parameters $U/I$ and
$Q/I$. $Q-U$ plots for all objects of our sample are available at
\textsl{http://www.iar.unlp.edu.ar/garra/garra-sdata.html}.

\section{Data analysis and results\label{s_dat}}

\subsection{Statistical analysis\label{s_sta}}

We quantitatively analyzed our data by computing a formal variability
indicator, following the criterion of \citet{K76}, which was used by several
other authors in variability studies \citep{A82, R94, ACR03}.  According to
this criterion, a source is classified as variable in an observing session
for the observable $S$ if the probability of exceeding the value
\begin{equation}
X^{2} = \sum_{i=1}^n{\epsilon_{i}^{-2}\, (S_i-\langle S
\rangle)^{2}}
\end{equation}
by chance is $< 0.1$ \%, and it is classified as non-variable if the
probability is $> 0.5$ \%. In this equation, $\epsilon_{i}$ is the error
corresponding to each measured value $S_i$, and $\langle S \rangle$ is the
mean value of $S$, given by:
\begin{equation}
\langle S \rangle = \frac {\sum_{i=1}^n{\epsilon_{i}^{-2}\, S_i}}
{\sum_{i=1}^n{\epsilon_{i}^{-2}}}.
\label{e_meanS}
\end{equation}
If the errors are random, $X^2$ should be distributed as $\chi^2$
with $n-1$ degrees of freedom, where $n$ is the number of points in
the distribution.

The other parameters that quantify the variability, in amplitude as
well as in timescale, are: the fluctuations index $\mu$,
\begin{equation}
\mu = 100 \frac{\sigma_s}{\langle S \rangle}\, \; {\%},
\end{equation}
where $\sigma_s$ is the standard deviation of one data set; the
fractional variability index of the source $FV$,
\begin{equation}
FV = {S_\mathrm{max} - S_\mathrm{min} \over S_\mathrm{max} + S_\mathrm{min}},
\end{equation}
where $S_\mathrm{max}$ and $S_\mathrm{min}$ are the maximum and
minimum values, respectively, for the polarization or the position
angle. Finally, the time interval $\Delta t$ between the extrema in the
polarization curve is defined as:
\begin{equation}
\Delta t = | t_{\rm max} - t_{\rm min} |,
\end{equation}
where $t_\mathrm{max}$ and $t_\mathrm{min}$ are the times when the
extreme points occur.\\

In Tables \ref{tab3} and \ref{tab4} we show the values of the
variability parameters for the linear polarization percentage and
the position angle for the RBLs and XBLs, respectively. Column 1
gives the name of the object, Col.~2 lists the observation dates,
Col.~3 shows the number of points for each night, Col.~4 gives the
total duration of the observation, Col.~5 gives the mean
polarization for the observing night using Equation~\ref{e_meanS},
Col.~6 shows the rms $\sigma_P$, Col.~7 gives the value for $FV$,
Col.~8 is $\Delta t$, Col.~9 shows the value of $\chi^2$, and
Col.~10 shows the variability class (V: if the source is variable,
NV: if it is not variable, and \emph{dubious} in the cases where
no definite decision could be reached using the above given
criteria). Cols.~11 to 16 give the same information for the
position angle.

\begin{sidewaystable*}
\begin{minipage}[t]{\textwidth}
\begin{tabular}{@{}lcrccccccccccccc}
\hline \noalign{\vskip 2pt} \hline \noalign{\smallskip}
&&&&
\multicolumn{6}{c}{Degree of polarization}&
\multicolumn{6}{c}{Position Angle}\\
&&&&
\multicolumn{6}{c}{\hrulefill}&
\multicolumn{6}{c}{\hrulefill}\\

Object & Date & $n$ & $\Delta t_{obs}$ & $<P>$ & $\sigma_{P}$ &
$F.V.$ & $\Delta t$ & $\chi ^{2}$ & $V/NV$ & $<\theta>$ &
$\sigma_{\theta}$ & $F.V.$ &
$\Delta t$ & $\chi ^{2}$ & $V/NV$ \\
  & [d/m/y] &  & [h] & [$\%$]~ &  &  & [h]  & &
  & [$^{\circ}$]~ &  & & [h]  & \\

\noalign {\smallskip} \hline\noalign{\vskip 2pt} \hline
\noalign{\smallskip}

0118-272 & 02/11/02 &  9 & 1.75 &8.17 & 1.091 & 0.24 & 0.9000 & 42.106 & $V$  & 139.43 & 1.539 & 0.026 & 0.7500 & 13.769 & $V$ \\
         & 03/11/02 & 14 & 2.22 &7.97 & 0.265 & 0.051 & 0.2482 &  7.632 & $NV$& 140.38 & 2.108 & 0.029 & 2.0145 & 39.788 & $V$ \\
0422+004 & 07/11/02 & 13 & 2.44 & 9.75 & 0.544 & 0.089 & 0.5667 & 53.461 & $V$ &  1.44 & 2.117 & 0.023 & 2.045 & 52.255 & $V$ \\
         & 08/11/02 & 17 & 3.44 & 11.52 & 0.739 & 0.132 & 1.9803 & 77.395 & $V$& 175.61 & 1.177 & 0.017 & 0.784 & 36.725 & $V$\\
0521-365 & 05/11/02 & 6 & 4.22 & 3.05 & 0.769 & 0.331 & 2.1387 & 61.38 & $V$ & 140.63 & 4.924 & 0.044 & 1.2391 & 48.146 & $V$ \\
         & 06/11/02 & 8 & 3.93 & 2.88 & 0.105 & 0.054 & 2.1621 & 3.583 & $NV$& 144.43 & 5.874 & 0.061 & 2.3018 & 82.220 & $V$ \\
0537-441 & 05/11/02 & 5 & 3.05 & 9.92 & 0.744 & 0.095 & 0.9363 & 5.539 & $Dubious$ & 6.93 & 2.120 & 0.377 & 0.936 & 6.594 & $V$ \\
         & 06/11/02 & 8 & 3.90 & 7.93 & 0.717 & 0.148 & 2.7658 & 16.412 & $V$ & 0.74 & 1.869 & 0.015 & 0.5347 & 9.716 & $Dubious$\\
0829+049 & 08/05/03 & 4 & 0.77 & 17.93 & 0.759 & 0.042 & 0.5137 & 9.108 & $V$ & 215.03 & 3.770 & 0.021 & 0.7740 & 97.224 & $V$ \\
1144-379 & 09/04/02 & 8 & 5.61 & 3.36 & 2.405 & 0.804 & 4.1942 & 15.849 & $V$& 117.48 & 24.269 & 0.258 & 1.138 & 37.343 & $V$ \\
         & 10/04/02 & 8 & 5.76 & 3.91 & 5.729 & 0.735 & 4.3768 & 85.500 & $V$& 136.64 & 7.426 & 0.088 & 4.0491 & 12.018 & $V$ \\
1510-089 & 11/04/02 & 9 & 5.20 & 3.33 & 4.177 & 0.910 & 2.5129 & 63.811 & $V$ & 79.25 & 44.994 & 0.734 & 0.5221 & 116.431 & $V$ \\
         & 15/04/04 & 12 & 3.85 &3.41 & 1.990 & 0.769 & 0.9415 & 16.299 & $V$  & 108.56 & 36.589 & 0.473 & 2.8886 & 47083.1 & $V$ \\
         & 16/04/04 & 8 & 3.23 & 3.35 & 1.738 & 0.727 & 0.7196 & 4.552 & $NV$ & 51.16 & 37.264 & 0.973 & 1.7916 & 46.579 & $V$ \\
1514+939 & 09/04/02 & 12 & 5.46 & 1.81 & 0.278 & 0.256 & 1.0489 & 46.857 & $V$& 155.52 & 10.666 & 0.086 & 2.9042 & 231.435 & $V$ \\
         & 10/05/03 & 11 & 3.08 & 2.91 & 0.442 & 0.258 & 1.5780 & 18.14 & $V$ & 157.62 & 4.060 & 0.039 & 2.7840 & 19.759 & $V$ \\
         & 18/04/04 & 6 & 1.49 & 3.33 & 0.329 & 0.113 & 0.626 & 12.254 & $V$ & 174.09 & 10.150 & 0.027 & 0.8911 & 13.444 & $V$ \\
1749+094 & 06/05/03 & 15 & 4.55 & 2.41 & 1.447 & 0.876 & 2.628 & 17.718 & $Dubious$& 81.771 & 62.561 & 0.644 & 1.9750 & 256.071 & $V$ \\
         & 07/05/03 & 14 & 4.66 & 4.66 & 2.489 & 0.739 & 0.579 & 30.725 & $V$ & 91.492 & 58.908 & 0.618 & 1.8922 & 530.175 & $V$ \\
2005-489 & 10/11/02 & 9 & 1.35 & 11.93 & 0.136 & 0.016 & 0.344 & 12.376 & $V$ & 93.560 & 0.262 & 0.003 & 1.0270 & 7.701 &$Dubious$\\

\hline
\noalign{\smallskip} \hline
\end{tabular}
\vfill
\caption{Variability results for the class of RBL}
\label{tab3}\centering
\end{minipage}
\end{sidewaystable*}

\begin{sidewaystable*}
\begin{minipage}[t]{\textwidth}
\begin{tabular}{@{}lcrccccccccccccc}
\hline \noalign{\vskip 2pt} \hline \noalign{\smallskip} &&&&
\multicolumn{6}{c}{Degree of polarization}&
\multicolumn{6}{c}{Position Angle}\\
&&&&\multicolumn{6}{c}{\hrulefill}&\multicolumn{6}{c}{\hrulefill}\\
Object & Date & $n$ & $\Delta t_{obs}$ & $<P>$ & $\sigma_{P}$ &
$F.V.$ & $\Delta t$ & $\chi ^{2}$ & $V/NV$ & $<\theta>$ &
$\sigma_{\theta}$ & $F.V.$ &
$\Delta t$ & $\chi ^{2}$ & $V/NV$ \\
  & [d/m/y] &  & [h] & [$\%$]~ &  &  & [h]  & &
  & [$^{\circ}$]~ &  & & [h]  & \\

\noalign {\smallskip} \hline\noalign{\vskip 2pt} \hline
\noalign{\smallskip}

0548-322 & 09/11/02 & 21 & 5.06 & 1.48 & 0.446 & 0.640 & 0.7730 & 44.687 & $V$& 38.24 & 16.447 & 0.646 & 3.4790 & 116.351 & $V$ \\
         & 10/11/02 & 15 & 1.35 & 1.31 & 0.383 & 0.469 & 1.3610 & 22.713 & $V$& 37.63 & 12.257 & 0.666 & 1.9260 & 41.939  & $V$ \\
0558-385 & 02/11/02 & 9 & 2.43 & 0.81 & 0.829 & 0.940 & 1.2323 & 40.135 & $V$ & 110.97 & 50.081 & 0.786 & 0.3138 & 149.150 & $V$\\
         & 03/11/02 & 11 & 2.68 & 0.73 & 0.966 & 0.881 & 0.8140 & 22.137 & $V$ & 89.17 & 40.305 & 0.648 & 0.8329 & 101.247 & $V$\\
1026-174 & 06/05/03 & 9 & 2.64 & 6.31 & 1.117 & 0.229 & 0.3229 & 6.436 & $Dubious$& 181.14 & 9.160 & 0.077 & 1.3460 & 15.106 & $V$ \\
         & 07/05/03 & 5 & 1.17 & 5.44 & 1.173 & 0.278 & 0.8190 & 5.615 & $Dubious$& 184.63 & 6.203 & 0.043 & 0.5590 & 4.792 & $Dubious$\\
1101-232 & 08/04/02 & 3 & 0.89 & 1.77 & 1.301 & 0.731 & 0.8853 & 6.729 & $V$ & 56.96 & 18.632 & 0.344 & 0.4461 & 0.955 & $Dubious$\\
         & 11/04/02 & 4 & 1.40 & 2.68 & 0.855 & 0.317 & 0.4100 & 3.761 & $Dubious$ & 73.02 & 9.081 & 0.132 & 0.3740 & 2.966 & $NV$ \\
         & 12/04/02 & 3 & 0.76 & 1.72 & 0.807 & 0.476 & 0.7585 & 1.805 & $Dubious$ & 70.34 & 8.846 & 0.104 & 0.3796 & 0.9800 & $NV$ \\
         & 09/05/03 & 3 & 2.04 & 8.38 & 4.713 & 0.376 & 0.3350 & 28.232 & $V$ & 8.84 & 11.634 & 0.948 & 0.3350 & 33.480 & $V$ \\
         & 15/04/04 & 8 & 2.76 & 1.68 & 0.981 & 0.622 & 0.3710 & 4.765 & $NV$ & 47.80 & 32.350 & 0.229 & 1.1411 & 23.578 & $V$ \\
1312-422 & 07/04/02 & 3 & 0.73 & 2.23 & 0.796 & 0.347 & 0.7760 & 0.699 & $NV$& 47.53 & 23.597 & 0.570 & 0.7760 & 2.258 & $V$ \\
         & 16/04/04 & 6 & 1.96 & 1.62 & 1.202 & 0.937 & 0.7929 & 6.476 & $Dubious$& 81.88 & 28.538 & 0.503 & 0.3613 & 6.897 & $Dubious$\\
1440+122 & 07/04/02 & 11 & 1.62 & 3.36 & 3.014 & 0.883 & 0.9375 & 39.004 & $V$& 90.65 & 30.748 & 0.655 & 0.5041 & 52.332 & $V$ \\
         & 08/04/02 & 11 & 4.92 & 2.40 & 2.223 & 0.769 & 2.2501 & 22.142 & $V$& 95.62 & 27.990 & 0.459 & 3.1650 & 32.290 & $V$ \\
1553+113 & 08/05/03 & 12 & 3.69 & 3.22 & 0.275 & 0.130 & 0.6070 & 58.190 & $V$ & 145.72 & 3.899 & 0.042 & 1.7940 & 169.885 & $V$ \\
         & 09/05/03 & 16 & 3.96 & 3.52 & 0.233 & 0.109 & 0.9913 & 43.744 & $V$ & 150.40 & 2.807 & 0.032 & 1.7616 & 122.118 & $V$ \\
2155-304 & 07/11/02 & 14 & 2.14 & 4.99 & 0.349 & 0.104 & 1.6264 & 63.442 & $V$ & 86.53 & 0.925 & 0.020 & 0.3102 & 14.979 & $Dubious$\\
         & 08/11/02 & 15 & 2.38 & 3.40 & 0.135 & 0.074 & 2.3804 & 72.483 & $V$ & 104.77 & 0.748 & 0.016 & 0.9676 & 23.177 & $V$ \\
         & 09/11/02 & 11 & 1.67 & 4.70 & 0.018 & 0.038 & 0.5227 & 56.893 & $V$ & 94.93 & 1.008 & 0.017 & 1.6744 & 74.863 & $V$ \\

\hline
\noalign{\smallskip} \hline
\end{tabular}
\vfill \caption{Variability results for the class of XBL}
\label{tab4}\centering
\end{minipage}
\end{sidewaystable*}

\subsection{Notes on individual sources \label{s_ind}}

We comment now briefly on the observed behaviour of each object in our sample.

\medskip

\noindent \textbf{RBLs:}
\paragraph{0118$-$272:} The linear optical polarization of this object was
measured by \citet{IT88} with a significantly high value
($P_V\sim17\%$). This was one of the reasons for its
classification as a blazar; a similar result was obtained by
\citet{MBB90}. We observed this blazar on two consecutive nights,
presenting variability on the first one, with the degree of
polarization rising from $P=5.77 \%$ to $P=9.41 \%$ in about one
hour. On the second night, the degree of polarization appears as
not variable but with a value of about $8 \%$. Meanwhile, the
position angle was variable with similar averages on both nights.

\paragraph{0422+004:} This blazar was reported to have quite high values of
linear optical polarization, between $7-22\%$ \citep{AS80}, presenting high
variability over long periods of time (months).  \citet{MBB90} observed the
source on two consecutive nights, detecting a decreasing degree of
polarization from the first night to the second one ($21.4\%$ -- $12.4\%$,
respectively). In Fig.~\ref{0422} we show the temporal evolution of the
linear polarization degree and position angle for this object, as an example
of the RBL class. This is the best sampled RBL in our campaign. It resulted
to be variable, both in $P$ and $\theta$, on the two nights. The degree of
polarization was quite high, mostly during the second night, reaching values
up to $\sim 13\%$.  The general trend was a smooth variation within each
night, with a higher mean value for the second night. These values are in
agreement with the published ones. The position angle was variable; however,
its mean value did not change significantly between both nights.

\paragraph{0521$-$365:} This southern blazar has been reported to have rapid
variability at radio frequencies \citep[see][]{RCV95}, as well as optical
flux microvariability \citep{RCCA02}. Its host was detected and classified
as a luminous giant elliptical by \citet{F94}.  During the present campaign,
the object experienced polarization variability on the first night, but the
variability is not significant on the second one. We think that the cause
for this could be the relatively large error bars for the degree of
polarization during that night, which could mask any variation. The typical
error was about $6\%$ of the measurement because of the weather
conditions. The mean value of the degree of linear polarization was almost
the same in the two nights, about $3\%$.  The position angle was clearly
variable during the two nights, with a small but clear rotation on the
second night: during the first hours, of $5~^{\circ}/$h in a clockwise
direction, and during the last hours, of $7.6~^{\circ}/$h in an
anticlockwise direction. Large rotations of the polarization angle have been
previously found at radio wavelengths by \citet{LMCR93} for this source.

\paragraph{0537$-$441:} This is another well-studied BL\,Lac object, which has
shown very high optical polarization during observations carried out by
Impey and Tapia in the 1980s \citep{IT88}. This object was extensively
monitored by \citet{RCC00aj,RCCA02}, presenting both behaviours, as
\emph{variable} and as \emph{non-variable} in its optical flux at different
epochs. We report here that during two consecutive nights in November 2002,
0537$-$441 presented a high degree of polarization.  On the first night the
object's variability appeared as \emph{dubious}, meanwhile on the second
night it was clearly $variable$, undergoing rapid fluctuations with typical
time scales of $\sim 1$ h and amplitudes of $\sim 1.4\%$. On the contrary,
$\theta$ was $variable$ on the first night and \emph{dubious} on the second
one, with a mean $\langle\theta\rangle \sim 0^{\circ}$.

\paragraph{0829+046:} Its first polarization observations at optical
wavelengths were presented by \citet{WW80}, who reported variability over a
few days time interval.  Unfortunately, we could follow this object just one
night. During this period, the object presented a very high degree of
polarization (up to $17\%$), increasing with time, and it was variable in
both the degree of polarization and position angle.  It is also interesting
to mention that this object was observed by \citet{GG04} at different radio
wavelengths in order to resolve the jet structure and they found that it is
one of the BL\,Lacs that presents evidence of emission on both sides of the
core: two symmetric jets were detected emerging from the core and both are
bended.

\paragraph{1144$-$379:} This object was classified as a blazar by
\citet{IT88}; they reported values for the degree of polarization
between $0.0\%$ and $9.4\%$. We observed the source in April 2002
and it was variable in both $P$ and $\theta$. During the last
night, the degree of polarization showed a peculiar behaviour: it
rose about $14\%$ in 4 hours, starting at $P=2.5\%$ and ending at
$P=16.5\%$. After ruling out all possible error sources (see
Sect.~\ref{s_obs}) we conclude that the cause of this peculiar
behaviour is intrinsic to the source.

\paragraph{1510$-$089:} This object was confirmed as a blazar by \citet{MS81}.
Previous measurements of its optical polarization degree, made in
$1980$, were all under $7.8\%$; however, \citet{MBB90} reported a
high value of $P=9.1\%$, in the $I$ band. We observed 1510$-$089
during three nights, one in April 2002, and two in April 2004. The
position angle was always variable. The degree of polarization was
variable in 2002 and on the first night in 2004, but not on the
second one. Its average value was about $3\%$ along all three
nights, but reaching values as high as $13.8\%$ during $2002$.
This BL\,Lac presented no microvariability during $1998$ and
$1999$ in its optical flux \citep{RCCA02}.

\paragraph{1514$-$241} (AP\,Lib): This is one of the objects which defined the
blazar class; it has presented values of optical polarization
between $2-7\%$ \citep {AS80}. \citet{MBB90} reported similar
values. We followed AP\,Lib on different occasions, resulting
always variable; however, the average degree of polarization was
quite low. During the last night in April 2004 the position angle
rotated in an anti-clockwise direction from $180.3^{\circ}$ to
$170.9^{\circ}$ with a speed of $10.5^{\circ}/$h.\footnote{Joining
this information with additional data obtained without filters,
the general trend in the position angle was a constant rotation
with sporadic direction reversals.}
\paragraph{1749+093:} This object has displayed dramatic polarization
variability at optical and infrared wavelengths \citep{BH86},
whereas no significant variations were detected later
\citep{MBB90}. Typical values for the degree of the optical
polarization are between $3-9\%$ \citep{KS90}. Our variability
data classify the source as \emph{dubious} during the first night
and variable on the second one, with a low mean value of the
degree of polarization, but reaching a maximum value of
$P_V=9.8\%$. The position angle was variable on both nights, but
this variability was probably not real, because when the modulus
of the polarization vector is small, the angle is ill defined,
thus preventing any real variability to be detected.
\paragraph{2005$-$489:} As far as we know, we are presenting here the
first optical polarization data for this BL\,Lac object. The
source was variable, with a relatively high degree of
polarization, during the only night when we could observe it. The
position angle variability was classified as \emph{dubious}, but
in fact, it remained almost constant around $93.7~^{\circ}$ with a
sigma of $0.2~^{\circ}$.

\medskip

\noindent \textbf{XBLs:}
\paragraph{0548$-$322:} \citet{AS80} had reported low levels in the degree of
polarization ($1.5-2\%$). Similar results were obtained during the campaign
undertaken by \citet{JS93}, when $P$ did not rise above $4\%$, with position
angle variable and showing a $0.5$ mag change in its optical flux. We
followed this typical BL\,Lac for two consecutive nights in November 2002,
with a good time resolution.  During both nights, the degree of optical
polarization appears to be variable but low, and the position angle was
variable, too.
\paragraph{0558$-$385:} We present the first polarization results for this
object. The average polarization is very low, $\langle P \rangle\sim 0.8-0.9$
\%. The object formally classifies as \emph{variable} but, since its
polarization is so low, the large fluctuations detected in the position
angle might be spurious.

\paragraph{1026$-$174:} There are no previous optical polarization data
for this blazar. We observed it in May 2003, when it displayed a
\emph{dubious} behaviour in its polarization degree.  However, during the
second night, some variation is present, unfortunately masked out by the
large error bars due to the weakness of the source. The highest and lowest
values of the optical polarization were $7.6\%$ and $3.9\%$,
respectively. The position angle was variable, showing no clear rotation
trend, with values around $185^{\circ}$.
\paragraph{1101$-$232:} This blazar has been reported to have quite low values
of optical polarization (the maximum detected was $2.7\%$), with evidences
of intrinsic variability \citep{JS93}. We have observed this source in 2002,
2003 and 2004. The behaviour displayed by the source went through different
stages (from $V$ to $NV$) along the different opportunities we had to
observe it. During the only night that we observed it in May 2003, $P_V$
rose to $14.7\%$, a very high value for this object, which had previously
presented lower polarization values.

\paragraph{1312$-$422}: This is another BL\,Lac with no previous data on
its optical polarization. We just followed it on one night in April 2002 and
another night in April 2004. The degree of polarization was quite low ($P_V
\lesssim 3.4\%$) and not variable on both occasions.

\paragraph{1440+122:} Published information about this object is scarce.
Recent radio observations \citep{GG04} revealed more details about
its structure, but no previous optical polarization data are
available. This blazar resulted to be variable in both $P_V$ and
$\theta$ during April 2002. $P_V$ reached as high a value as
$\approx 8.3\%$, with a peculiar behaviour during the first night
we followed it. After discarding any possible error sources (see
Sect.~\ref{s_obs}), the trend in the polarization degree is an
interesting one, rising from $P_V=1.3\%$ to $P_V=8.3\%$ during the
first $1.8$~h, then going down to $P_V=0.5\%$ in about $1$ h, and
finally rising again up to $P_V=7.5\%$.  Meanwhile, the mean
polarization was not too high ($\langle P_V \rangle = 3.4\%$). On
the second night, $P_V$ was variable, but with no peculiar trend.
The position angle did not follow the variations in $P_V$;
however, it was always variable, with values around $\theta \simeq
95^{\circ}$.

\paragraph{1553+133:} The degree of polarization of this BL\,Lac object was
variable during our observations, with a flickering behaviour on
the first night. A qualitatively similar flickering has been
detected before at radio wavelengths in extragalactic radio
sources \citep{H84, RBC95}. In the optical polarization the origin
of the rapid flickering must be intrinsic to the source, probably
related with turbulence in the magnetic field of the inner jet.
The upper limit of the polarization was $4.2\%$ for this object.

\paragraph{2155-304:} This is a very well studied BL\,Lac object,
known for its short variability time scales at optical to X-ray
wavelengths \citep{JS93}. \citet{AS80} reported values for the
degree of optical polarization between $3-7\%$. In the mid 1980s,
\citet{BH86} detected polarization percentage variations and a
clockwise rotation of the position angle, although on a $\sim 48$
h time scale. Four years later, \citet{MBB90} measured a higher
than usual polarization ($P_V \approx 10\%$). The extensive
monitoring made by \citet{SHAS92} in optical polarization and
photometry revealed clear variations over long time scales; the
authors also reported a $2\%-3\%$ variation in $P_{V}$ and as much
as $25^{\circ}$ in $\theta$, in $24$~h. They also found a mild
wavelength dependent polarization and a more rapid variation in
the optical polarization than in the total optical flux. In a more
recent work \citet{TDP01} reported a multiband monitoring of the
optical polarization searching for intranight and also long term
variability. This campaign was made using the 2.15~m telescope at
CASLEO equipped with the Photopolarimeter of the Turin Observatory
and the total observing time was $\sim 47$~h along four different
periods in 1998 and 1999. The authors reported a $P_{V}$ lower
than $7\%$ with small amplitude intranight variations, $\sim
1.3\%$ in $P$ and $\sim 7^{\circ}$ in $\theta$, but no statistical
analysis of this behaviour was reported. They also found a low
wavelength dependence in both the linear polarization and the
position angle.

Being a relatively bright object, we were able to use short
exposure times, hence getting well-sampled time curves. These are
shown, as an example for the XBL class, in Fig.\ref{2155}.
Significant variations both in $P_V$ and $\theta$ are clearly seen
on each of the three consecutive nights that we followed the
source, with a moderately high mean polarization $\langle P_V
\rangle \sim \mathrm{5}\%$ and with a position angle varying from
$\langle \theta \rangle = 87 ^{\circ}$ to $\langle \theta \rangle
= 105 ^{\circ}$. During the first night, $P_{V}$ raised as high as
$5.7\%$ at the beginning of the night and then went down, ending
at $4.7\%$. Apparently, this decreasing trend continued during the
day hours, because at the beginning of the second night, $P_{V}$
started at $3.2\%$, going up for the rest of the night and
presenting an inverse behaviour during the last night. With
respect to the position angle, the variation was clear and
presented a fast rotation during the third night. The angle
rotated in an anti-clockwise direction from $96.7^{\circ}$ to
$93.5^{\circ}$ with a speed of $1.9 ^{\circ}/{\rm h}$.

\section{Discussion \label{s_disc}}

In order to characterize the two different classes of objects under study
here, we analyzed the behaviour of the sources from a statistical point
of view. First, we calculated the duty cycles (DC) for the sources of a
given class. This quantity can be estimated, following
\citet{RCC99,RCCA02}, as
\begin{equation}
DC = 100 \frac{\sum_{i=1}^n {N_i , (1/\Delta t_i)}}{\sum_{i=1}^n
{(1/\Delta t_i)}} {\%},
\end{equation}
where $\Delta t_i = \Delta t_{i,\mathrm{obs}}(1+z)^{-1}$ is the duration,
corrected by the corresponding redshift, of the $i$-th data set of the
quantity and class under study; $N_i$ is the weight (equal to 1 if the
source was classified as $V$, or 0 if the source was NV or
\emph{dubious}). Because we weighted \emph{dubious} cases with 0, the DCs
calculated are actually lower limits. The corresponding DCs for the degree
of polarization ($P$) and position angle ($\theta$) for both classes of
objects (RBLs and XBLs) are: $DC(P,\mathrm{RBL})=77.01 \%$,
$DC(\theta,\mathrm{RBL})=87.25\%$, and $DC(P,\mathrm{XBL})=51.23 \%$,
$DC(\theta,\mathrm{XBL})=55.15\%$. So, the RBLs appear to constitute the
most variable class. A similar behaviour has been found when only optical
flux variations were considered by \citet{RCCA02}, with duty cycles
$DC=71.5\%$ and $DC=27.9\%$ for the RBLs and XBLs, respectively. The
photometric microvariability of XBLs seems to be systematically lower,
nonetheless, than their polarization microvariability.

\begin{figure*}
\includegraphics[width=0.45\hsize]{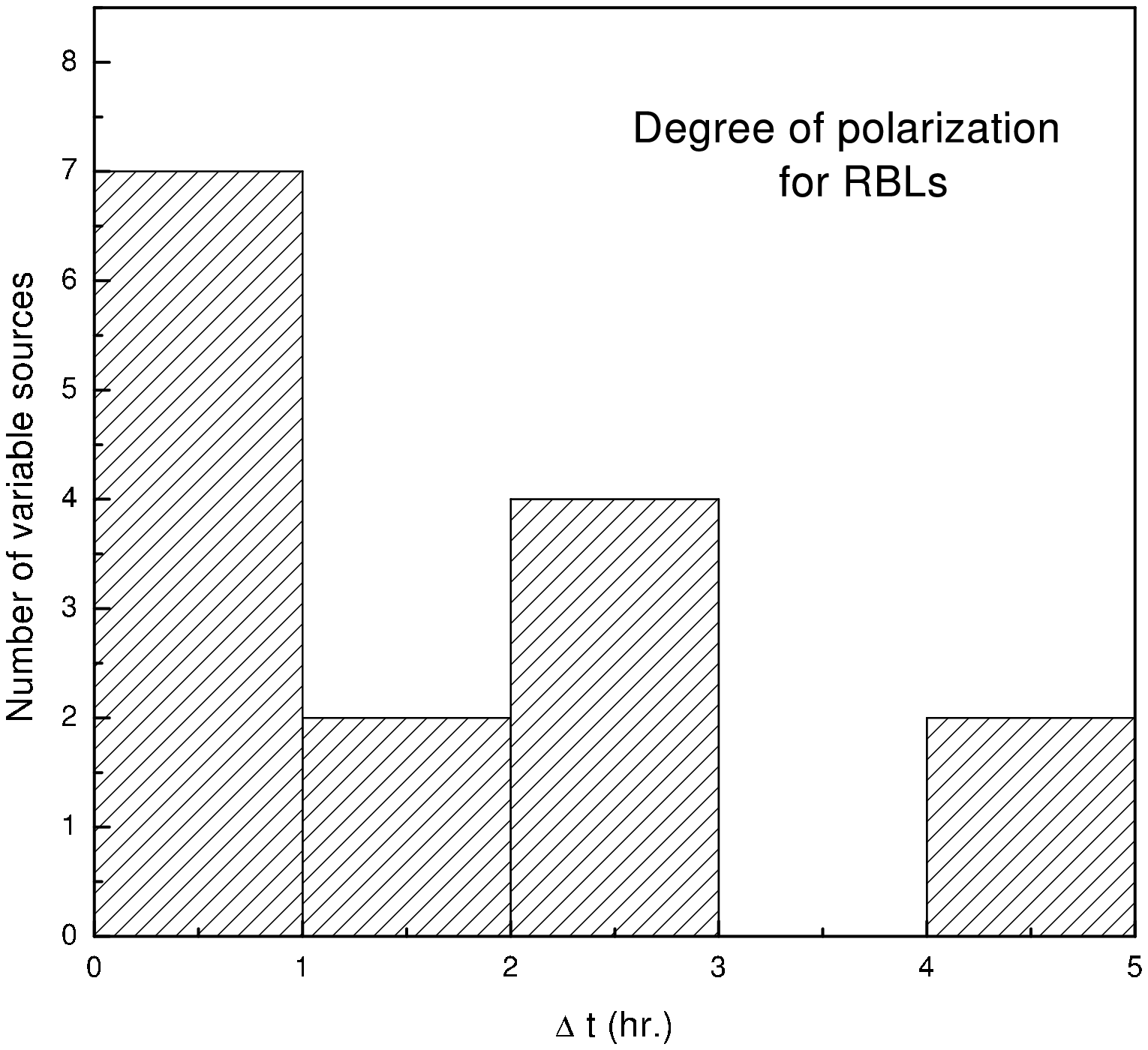} \qquad
\includegraphics[width=0.45\hsize]{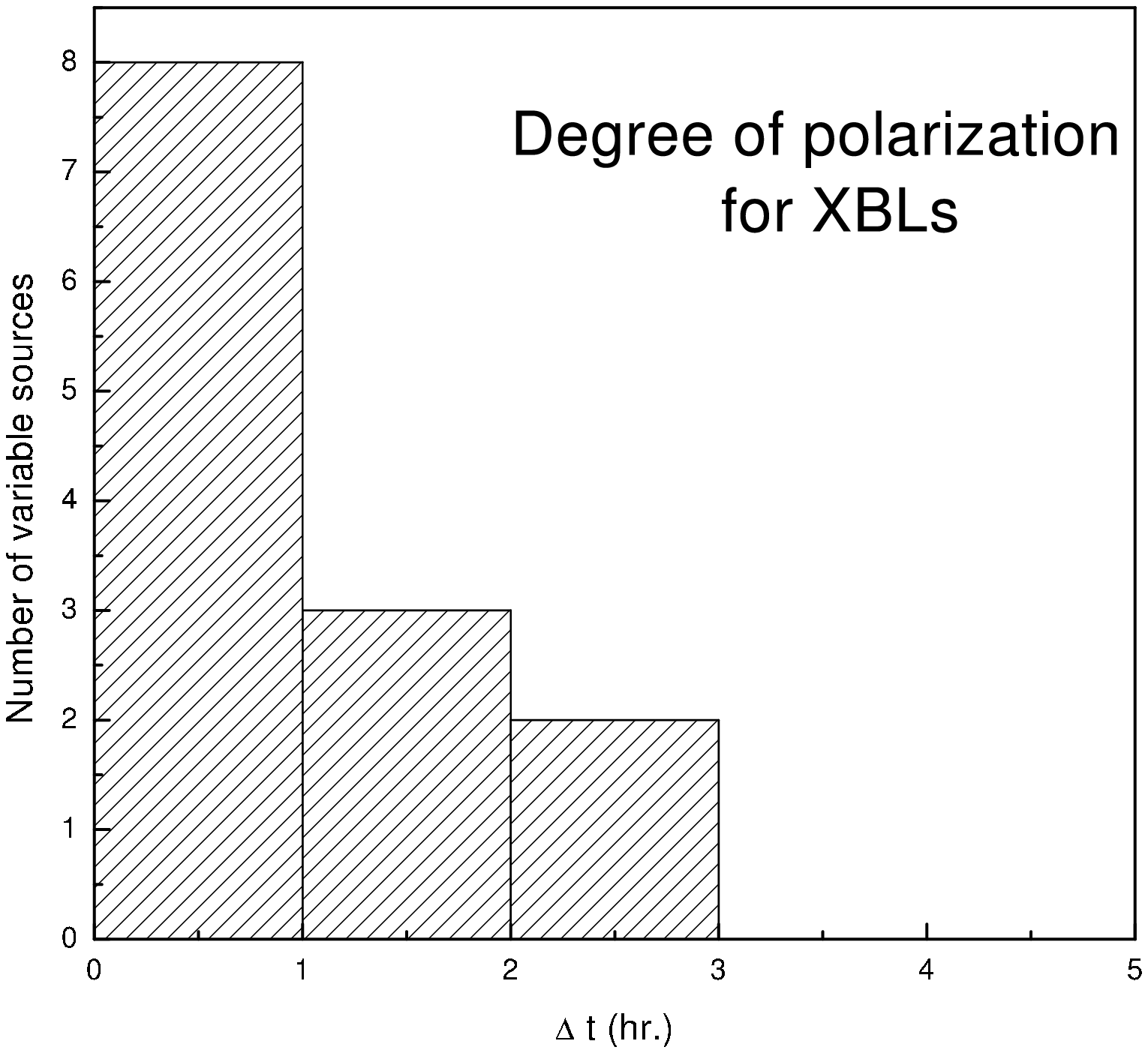}
\caption{Histogram with the distribution of variable RBLs and
XBLs against the variability timescales (degree of polarization)}
\label{pol_a}
\end{figure*}

\begin{figure*}
\includegraphics[width=0.45\hsize]{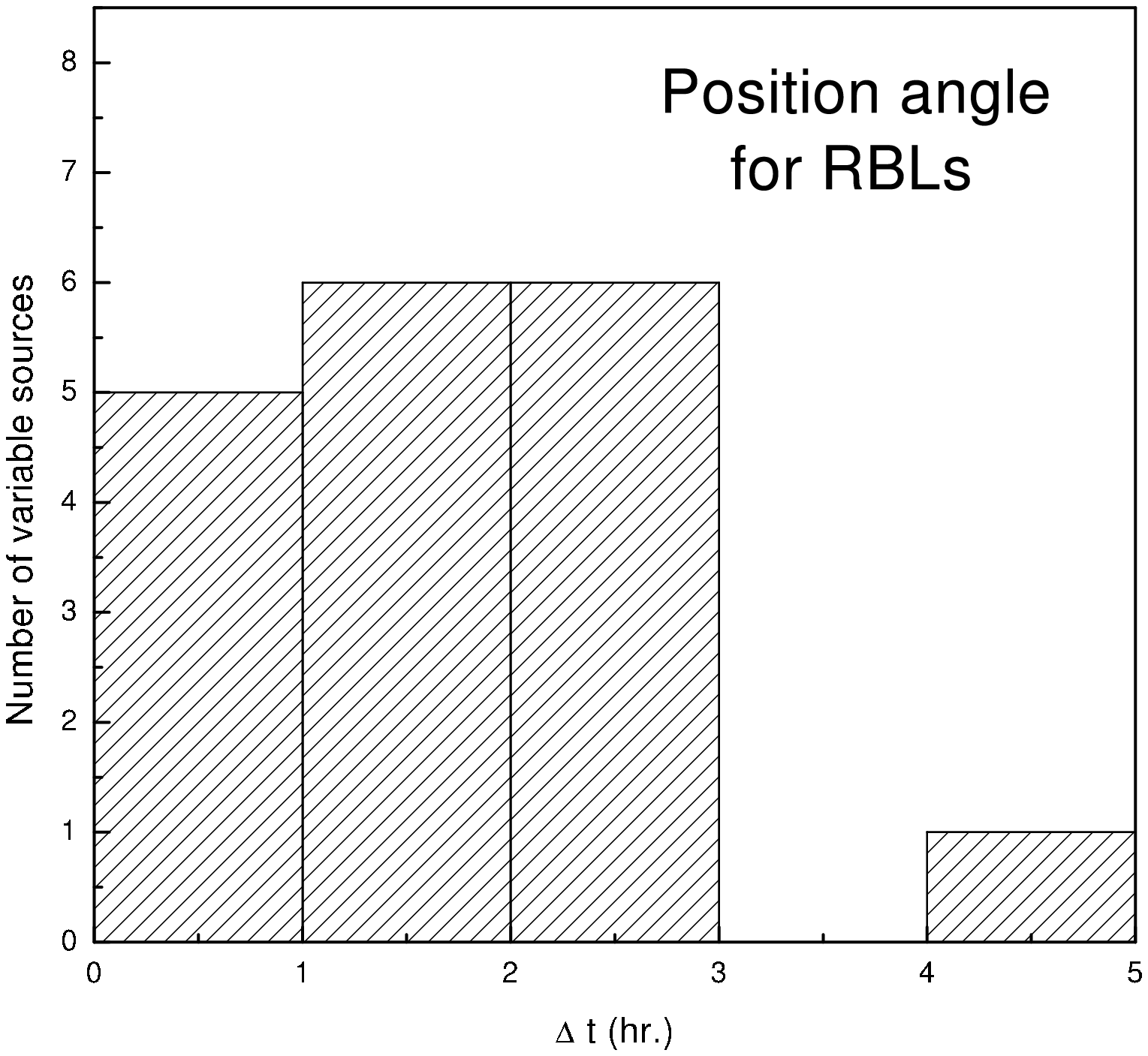} \qquad
\includegraphics[width=0.45\hsize]{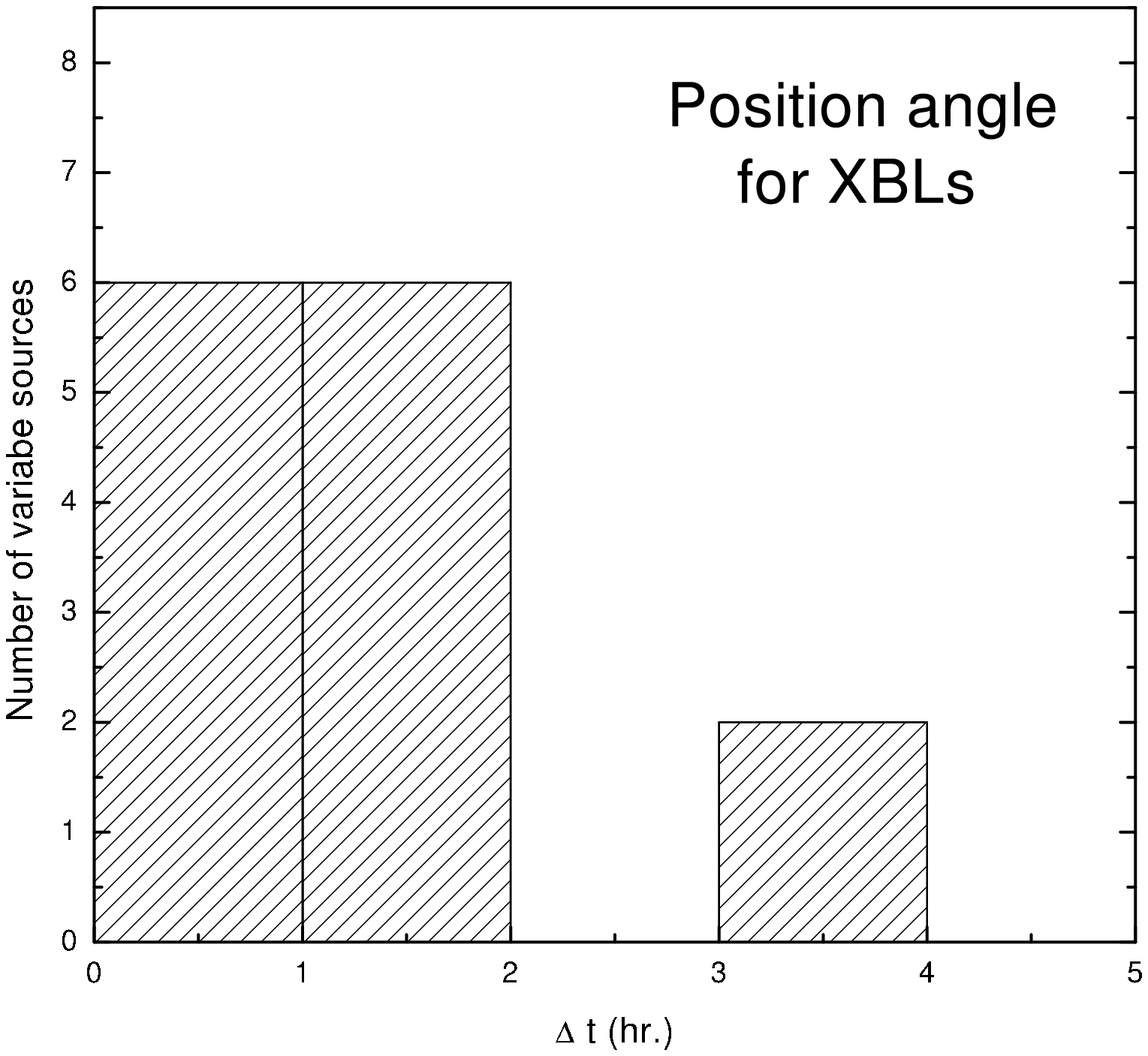}
\caption{Idem Fig.~\ref{pol_a} (position angle)} \label{ang_a}
\end{figure*}

\begin{figure*}[t]
\includegraphics[width=0.45\hsize]{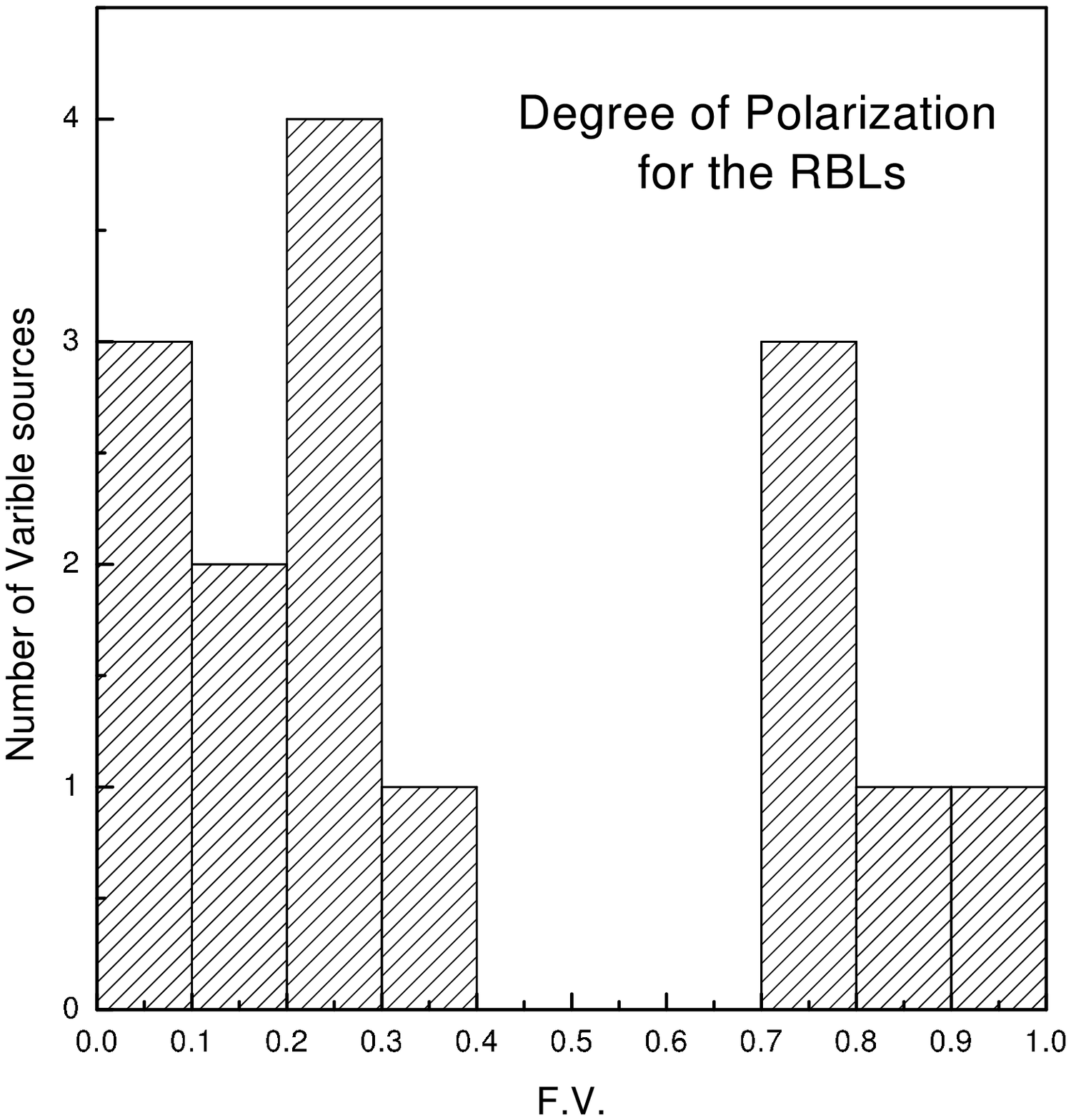} \qquad
\includegraphics[width=0.45\hsize]{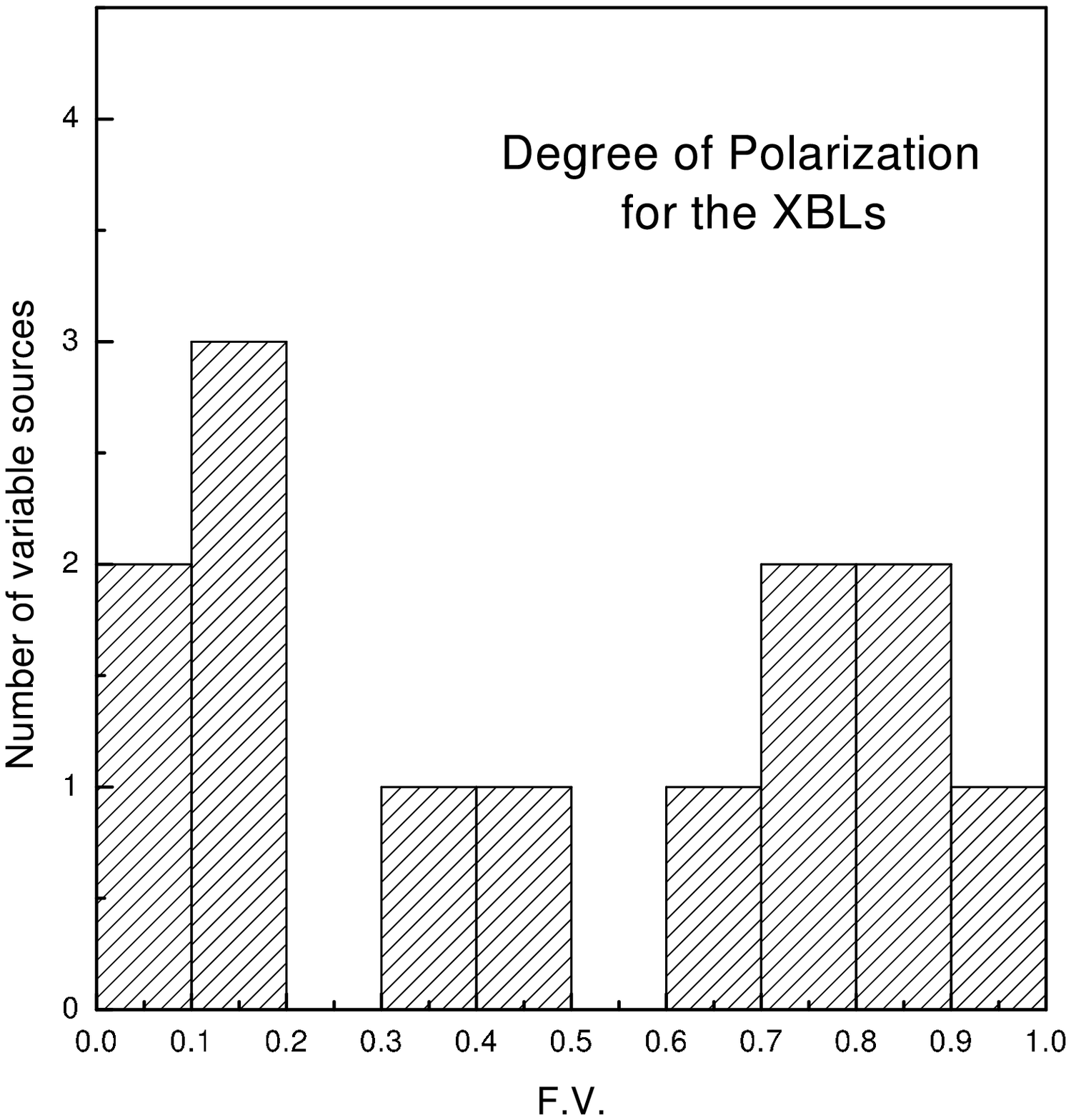}
\caption{Histograms with the distribution of variable RBLs and
XBLs against the fractional variability index (degree of
polarization)} \label{pol_b}
\end{figure*}

\begin{figure*}[t]
\includegraphics[width=0.45\hsize]{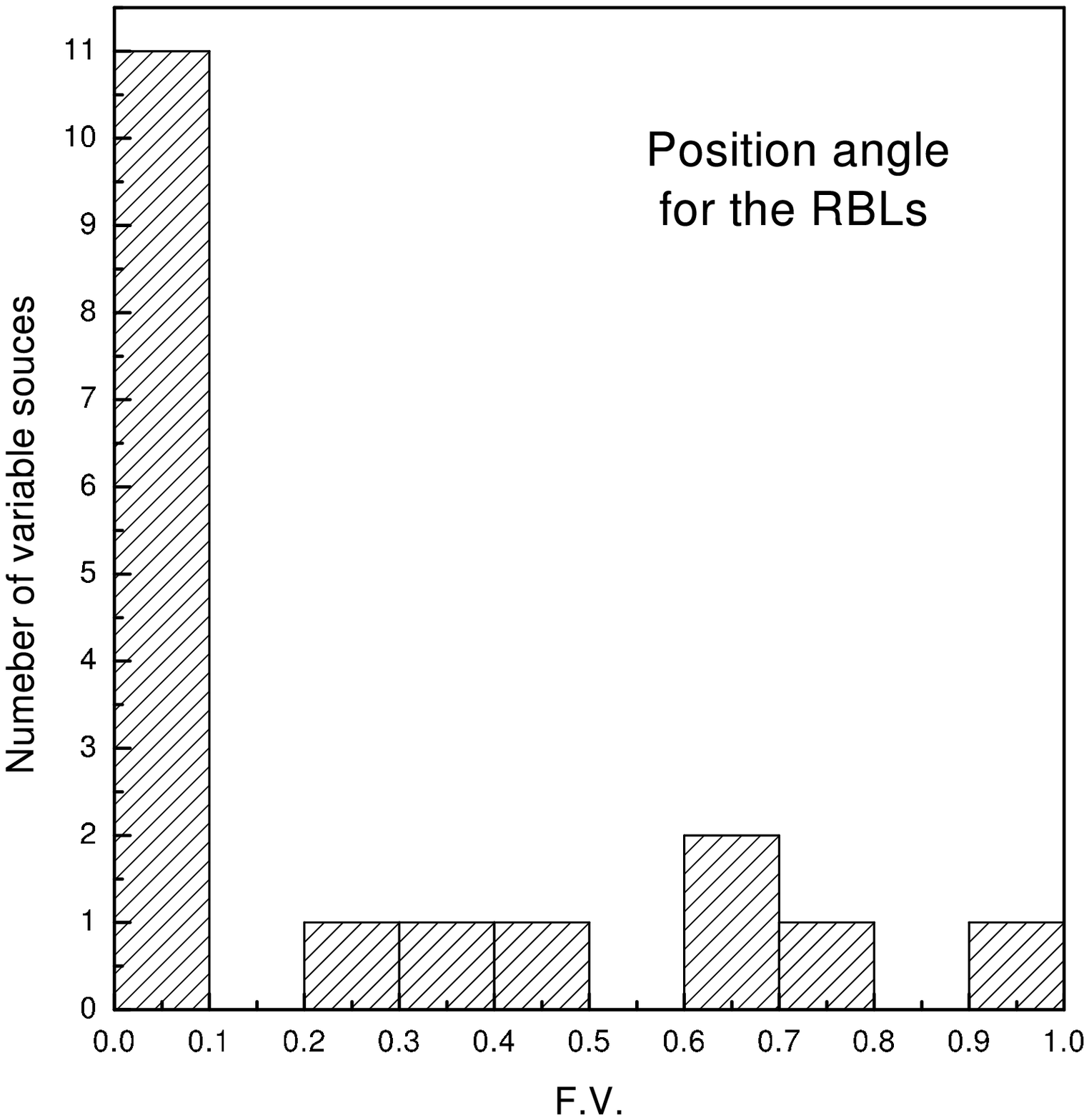} \qquad
\includegraphics[width=0.45\hsize]{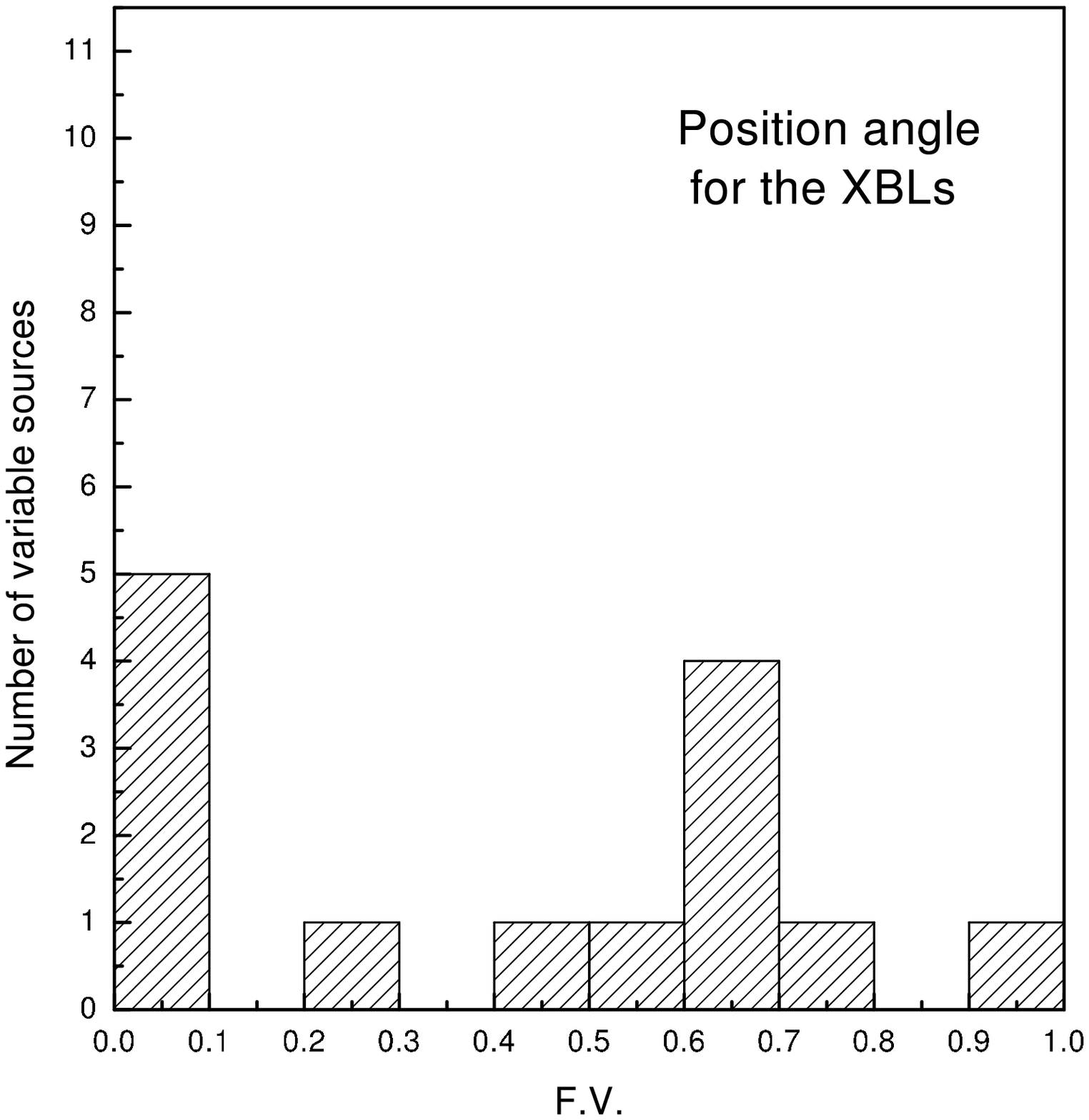}
\caption{Idem Fig.~\ref{pol_b} (position angle)} \label{ang_b}
\end{figure*}

\begin{figure*}[t]
\includegraphics[width=0.45\hsize]{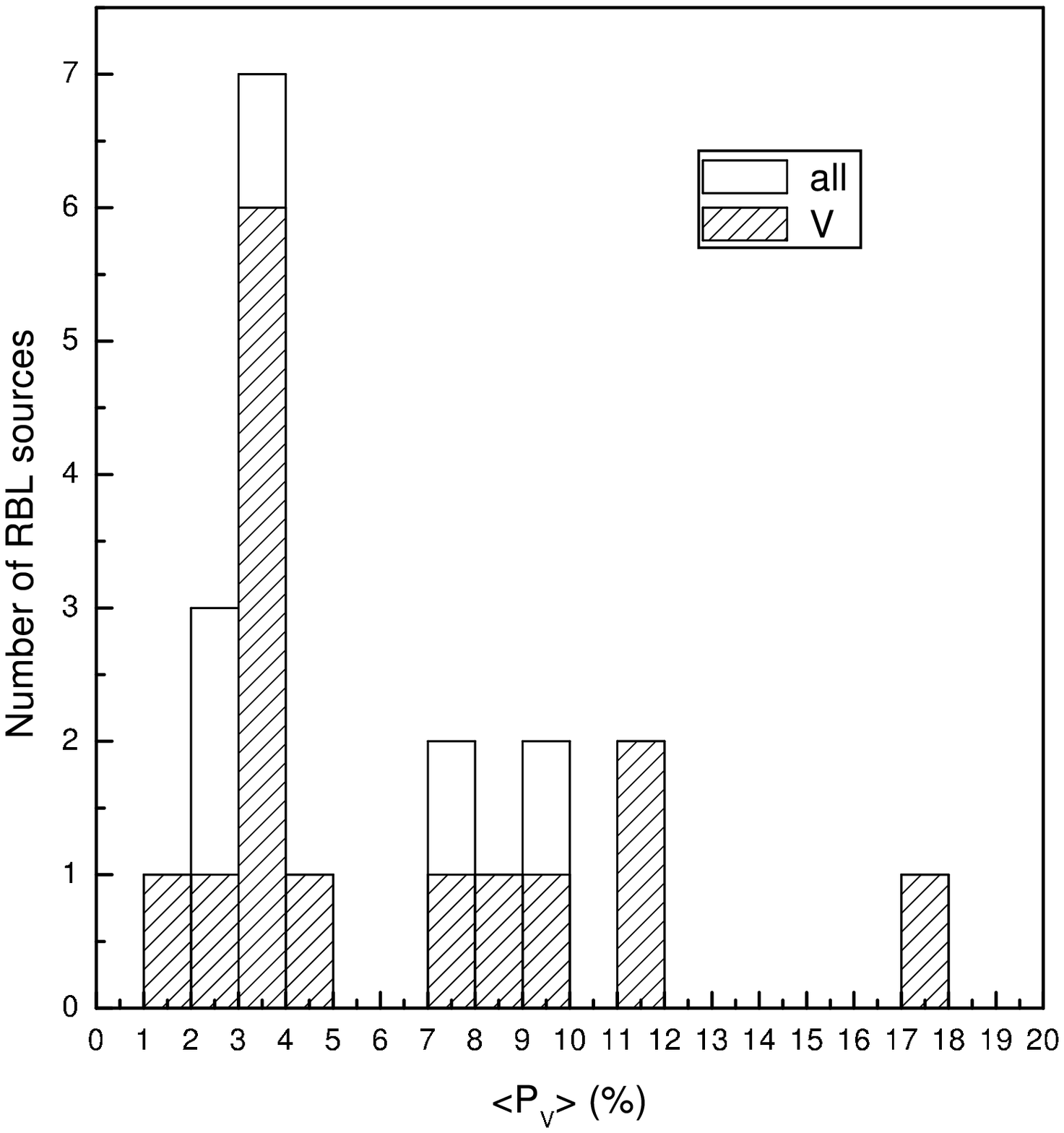} \qquad
\includegraphics[width=0.45\hsize]{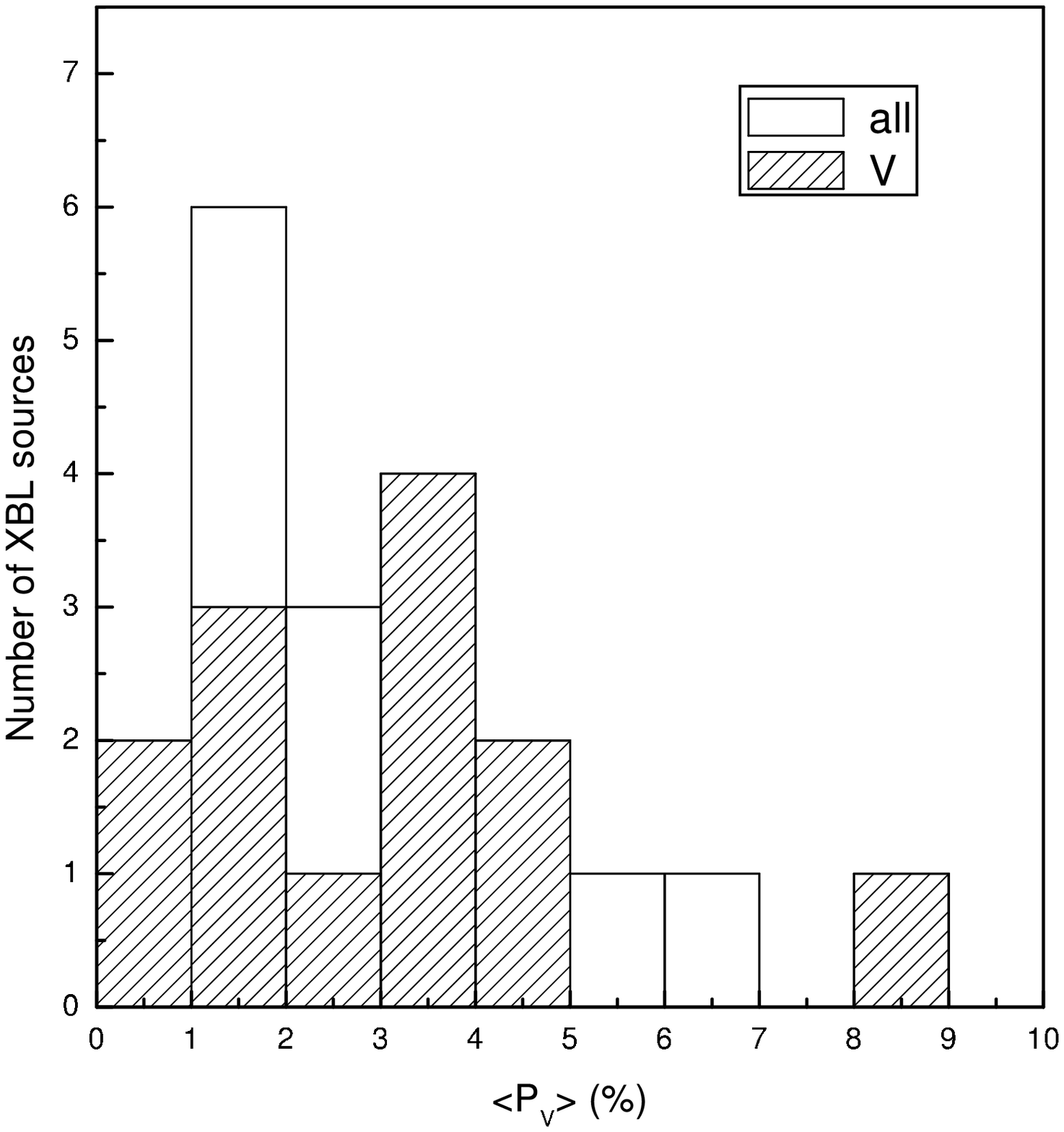}
\caption{Histograms with the distributions of all observed sources
(left: RBLs; right: XBLs) against the mean value of the
degree of optical polarization. \label{pol_prom1}}
\end{figure*}

A complementary view can be obtained by plotting the histograms of
the distributions of the sources that resulted to be variable.
Figures~\ref{pol_a} and \ref{ang_a} show the number of sources
classified as \emph{variable} against the time scales for the
variation (Col.~8 and 14 in Tables \ref{tab3} and \ref{tab4}), for
the degree of polarization and position angle, respectively.

It can be seen that the RBLs have a wider and flatter distribution in the
degree of polarization than the XBLs; this is an indicator that, when an
XBLs is variable, its variation timescale is shorter than that of the RBLs
(typically $\Delta t \lesssim 1$ h). The histograms corresponding to the
position angles present no significant differences between the two classes.

Similar histograms were made showing the distributions of variable
sources against the fractional variability index (Col.~ 7 and 13
in the same tables). This is shown in Figs.~\ref{pol_b} and
\ref{ang_b} for the same parameters as before. Here, the RBLs
appear to have two peaks, one for $P \approx 0.2 - 0.3 \%$, and
the other around $P \approx 0.8 - 0.9$. On the contrary, the XBLs
have a more uniform distribution.

Concerning the position angle, the variation of the RBLs appears to be more
frequent for the lowest $FV$ values. Again, the XBLs seem to have a more
uniform distribution.

Since we present here a significant number of sources belonging to two
different sub-types of BL\,Lac objects, it is also interesting to compare
the average degree of optical polarization between them. Our results confirm
the previous ones reported by \citet{FCZL97}: the XBLs generally have lower
optical polarization than the RBLs; this kind of behaviour can be seen in
the histograms drawn in Figure~\ref{pol_prom1}. A Kolmogorov-Smirnov test
shows that both data sets are most probably taken from different parent
distributions, although the significance level is only moderately high
(95\%). As an additional information, we also include the distribution of V
and NV or \emph{dubious} cases for both classes of BL\,Lacs studied here.

In general, since XBLs have the synchrotron peak of their spectral
energy distribution at X-ray energies, we could expect that these
objects have, on average, either higher magnetic fields {\em or}
more energetic particles than RBLs, which peak at radio-IR
wavelengths. The fact that they have on average less polarization
and that this polarization is less variable than what is found for
RBLs seems to support the second possibility, i.e., the particles
in their jets are systematically more energetic than in RBLs. On
the contrary, as noticed by \citet{FCZL97}, the RBLs seem to have
larger macroscopic relativistic motions, hence displaying higher
duty cycles for rapid variability, which is probably associated
with relativistic shocks in the jets. Their magnetic fields seem
to be also systematically stronger than in XBLs, as indicated by
the higher degree of linear polarization. This leads to a simple
picture where XBLs have particles with high microscopic Lorentz
factors that cool radiating high-energy synchrotron emission
whereas RBLs have less energetic particles but higher macroscopic
bulk motions and stronger fields, hence presenting higher
variability. Alternatively, XBL could have similar magnetic
fields, but less homogeneous, hence less degree of linear
polarization. The origin of the rapid microvariability seems to be
associated with relativistic shocks in any case \citep[e.g.][and
references therein]{RCV95}.

\section{Conclusions \label{s_conc}}

We have monitored 8 XBLs and 10 RBLs, looking for intranight
variability in the optical polarization. We have found high duty
cycles for both the degree of linear polarization and the
polarization angle of RBLs. The average polarization is also
stronger than for XBLs. XBLs, although displaying a lower level of
polarimetric microvariability, are also significantly variable,
with duty cycles of $\sim 50$ \%, higher than what is observed
from purely photometric observations. We speculate, on the basis
of our findings, that the stronger synchrotron losses presented by
XBLs might be due to systematically higher microscopic Lorentz
factors for the particles in the jets, rather than to stronger
magnetic fields. However, it should be noted that further
observations of objects (with each object monitored on several
nights) are needed to establish this conclusion firmly.

\begin{acknowledgements}
This work has been supported by CONICET and ANPCyT (through Grant PICT
03-13291). We thank the staff of CASLEO observatory for valuable help during
the observations.  We also thank the constructive suggestions made by the
anonymous referee. This research made use of the NASA Extragalactic Data
system.
\end{acknowledgements}


\begin{thebibliography}{39}
\expandafter\ifx\csname natexlab\endcsname\relax\def\natexlab#1{#1}\fi

\bibitem[{{Altschuler}(1982)}]{A82}
{Altschuler}, D.~R. 1982, \aj, 87, 387

\bibitem[{{Andruchow} {et~al.}(2003){Andruchow}, {Cellone}, {Romero},
  {Dominici}, \& {Abraham}}]{ACR03}
{Andruchow}, I., {Cellone}, S.~A., {Romero}, G.~E., {Dominici}, T.~P., \&
  {Abraham}, Z. 2003, \aap, 409, 857

\bibitem[{{Angel} \& {Stockman}(1980)}]{AS80}
{Angel}, J.~R.~P. \& {Stockman}, H.~S. 1980, \araa, 18, 321

\bibitem[{{Brindle} {et~al.}(1986){Brindle}, {Hough}, {Bailey}, {Axon}, \&
  {Hyland}}]{BH86}
{Brindle}, C., {Hough}, J.~H., {Bailey}, J.~A., {Axon}, D.~J., \& {Hyland},
  A.~R. 1986, \mnras, 221, 739

\bibitem[{{Falomo}(1994)}]{F94}
{Falomo}, R. 1994, The Messenger, 77, 49

\bibitem[{{Fan} {et~al.}(1997){Fan}, {Cheng}, {Zhang}, \& {Liu}}]{FCZL97}
{Fan}, J.~H., {Cheng}, K.~S., {Zhang}, L., \& {Liu}, C.~H. 1997, \aap, 327, 947

\bibitem[{{Giroletti} {et~al.}(2004){Giroletti}, {Giovannini}, {Taylor}, \&
  {Falomo}}]{GG04}
{Giroletti}, M., {Giovannini}, G., {Taylor}, G.~B., \& {Falomo}, R. 2004, \apj,
  613, 752

\bibitem[{{Heeschen}(1984)}]{H84}
{Heeschen}, D.~S. 1984, \aj, 89, 1111

\bibitem[{{Heidt} \& {Wagner}(1996)}]{HW96}
{Heidt}, J. \& {Wagner}, S.~J. 1996, \aap, 305, 42

\bibitem[{{Heidt} \& {Wagner}(1998)}]{HW98}
{Heidt}, J. \& {Wagner}, S.~J. 1998, \aap, 329, 853

\bibitem[{{Hough}(1996)}]{H96}
{Hough}, J.~H. 1996, in ASP Conf.\ Ser., Vol.~97, 569

\bibitem[{{Impey} \& {Tapia}(1988)}]{IT88}
{Impey}, C.~D. \& {Tapia}, S. 1988, \apj, 333, 666

\bibitem[{{Impey} \& {Tapia}(1990)}]{IT90}
{Impey}, C.~D. \& {Tapia}, S. 1990, \apj, 354, 124

\bibitem[{{Jang} \& {Miller}(1997)}]{JM97}
{Jang}, M. \& {Miller}, H.~R. 1997, \aj, 114, 565

\bibitem[{{Jannuzi} {et~al.}(1993){Jannuzi}, {Smith}, \& {Elston}}]{JS93}
{Jannuzi}, B.~T., {Smith}, P.~S., \& {Elston}, R. 1993, \apjs, 85, 265

\bibitem[{{Kesteven} {et~al.}(1976){Kesteven}, {Bridle}, \& {Brandie}}]{K76}
{Kesteven}, M.~J.~L., {Bridle}, A.~H., \& {Brandie}, G.~W. 1976, \aj, 81, 919

\bibitem[{{K\"uhr} \& {Schmidt}(1990)}]{KS90}
{K\"uhr}, H. \& {Schmidt}, G.~D. 1990, \aj, 99, 1

\bibitem[{{Luna} {et~al.}(1993){Luna}, {Martinez}, {Combi}, \&
  {Romero}}]{LMCR93}
{Luna}, H.~G., {Martinez}, R.~E., {Combi}, J.~A., \& {Romero}, G.~E. 1993,
  \aap, 269, 77

\bibitem[{{Magalh\~aes} {et~al.}(1984){Magalh\~aes}, {Benedetti}, \&
  {Roland}}]{M84}
{Magalh\~aes}, A.~M., {Benedetti}, E., \& {Roland}, E.~H. 1984, \pasp, 96, 383

\bibitem[{Mart\'{\i}nez {et~al.}(1990)Mart\'{\i}nez, Aballay, Mar\'un, \&
  Ruartes}]{Martinez}
Mart\'{\i}nez, E., Aballay, J.~L., Mar\'un, A., \& Ruartes, H. 1990, Bol.\
  Asoc.\ Arg.\ de Astronom\'{\i}a, 36, 342

\bibitem[{{Mead} {et~al.}(1990){Mead}, {Ballard}, {Brand}, {Hough}, {Brindle},
  \& {Bailey}}]{MBB90}
{Mead}, A.~R.~G., {Ballard}, K.~R., {Brand}, P.~W.~J.~L., {et~al.} 1990, \aaps,
  83, 183

\bibitem[{{Miller} {et~al.}(1989){Miller}, {Carini}, \& {Goodrich}}]{MCG89}
{Miller}, H.~R., {Carini}, M.~T., \& {Goodrich}, B.~D. 1989, \nat, 337, 627

\bibitem[{{Moore} \& {Stockman}(1981)}]{MS81}
{Moore}, R.~L. \& {Stockman}, H.~S. 1981, \apj, 243, 60

\bibitem[{{Padovani} \& {Giommi}(1995)}]{PG95}
{Padovani}, P. \& {Giommi}, P. 1995, \mnras, 277, 1477

\bibitem[{{Racine}(1970)}]{R70}
{Racine}, R. 1970, \apjl, 159, L99

\bibitem[{{Romero} {et~al.}(1994){Romero}, {Combi}, \& {Colomb}}]{R94}
{Romero}, G.~E., {Combi}, J.~A., \& {Colomb}, F.~R. 1994, \aap, 288, 731

\bibitem[{{Romero} {et~al.}(1995{\natexlab{a}}){Romero}, {Benaglia}, \&
  {Combi}}]{RBC95}
{Romero}, G.~E., {Benaglia}, P., \& {Combi}, J.~A. 1995{\natexlab{a}}, \aap,
  301, 33

\bibitem[{{Romero} {et~al.}(1995{\natexlab{b}}){Romero}, {Combi}, \&
  {Vucetich}}]{RCV95}
{Romero}, G.~E., {Combi}, J.~A., \& {Vucetich}, H. 1995{\natexlab{b}}, \apss,
  225, 183

\bibitem[{{Romero} {et~al.}(1999){Romero}, {Cellone}, \& {Combi}}]{RCC99}
{Romero}, G.~E., {Cellone}, S.~A., \& {Combi}, J.~A. 1999, \aaps, 135, 477

\bibitem[{{Romero} {et~al.}(2000){Romero}, {Cellone}, \& {Combi}}]{RCC00aj}
{Romero}, G.~E., {Cellone}, S.~A., \& {Combi}, J.~A. 2000, \aj, 120, 1192

\bibitem[{{Romero} {et~al.}(2002){Romero}, {Cellone}, {Combi}, \&
  {Andruchow}}]{RCCA02}
{Romero}, G.~E., {Cellone}, S.~A., {Combi}, J.~A., \& {Andruchow}, I. 2002,
  \aap, 390, 431

\bibitem[{{Scarpa} \& {Falomo}(1997)}]{SF97}
{Scarpa}, R. \& {Falomo}, R. 1997, \aap, 325, 109

\bibitem[{{Smith} {et~al.}(1992){Smith}, {Hall}, {Allen}, \& {Sitko}}]{SHAS92}
{Smith}, P.~S., {Hall}, P.~B., {Allen}, R.~G., \& {Sitko}, M.~L. 1992, \apj,
  400, 115

\bibitem[{{Stalin} {et~al.}(2004){Stalin}, {Gopal-Krishna}, {Sagar}, \&
  {Wiita}}]{SGSW04}
{Stalin}, C.~S., {Gopal-Krishna}, G., {Sagar}, R., \& {Wiita}, P.~J. 2004,
  \mnras, 350, 175

\bibitem[{{Stalin} {et~al.}(2005){Stalin}, {Gupta}, {Gopal-Krishna}, {Wiita},
  \& {Sagar}}]{SGG05}
{Stalin}, C.~S., {Gupta}, A.~C., {Gopal-Krishna}, G., {Wiita}, P.~J., \&
  {Sagar}, R. 2005, \mnras, 356, 607

\bibitem[{{Tommasi} {et~al.}(2001){Tommasi}, {D{\'{\i}}az}, {Palazzi}, {Pian},
  {Poretti}, {Scaltriti}, \& {Treves}}]{TDP01}
{Tommasi}, L., {D{\'{\i}}az}, R., {Palazzi}, E., {et~al.} 2001, \apjs, 132, 73

\bibitem[{{Turnshek} {et~al.}(1990){Turnshek}, {Bohlin}, {Williamson}, {Lupie},
  {Koornneef}, \& {Morgan}}]{T90}
{Turnshek}, D.~A., {Bohlin}, R.~C., {Williamson}, R.~L., {et~al.} 1990, \aj,
  99, 1243

\bibitem[{{V\'eron-Cetty} \& {V\'eron}(1998)}]{V98}
{V\'eron-Cetty}, M.-P. \& {V\'eron}, P. 1998, ESO Scientific Report
Series, Vol.~18, {A Catalogue of quasars and active nuclei}, 8th
edn. (Garching: European Southern Observatory)

\bibitem[{{Wills} {et~al.}(1980){Wills}, {Wills}, {Breger}, \& {Hsu}}]{WW80}
{Wills}, D., {Wills}, B.~J., {Breger}, M., \& {Hsu}, J.-C. 1980, \aj, 85, 1555

\end{thebibliography}
\end{document}